\documentclass[useAMS,usenatbib]{mn2e}

\usepackage{aas_macros}    
\usepackage{amsmath}       
\usepackage{amssymb}
\usepackage{wasysym}	   
\usepackage{mathrsfs}      
\usepackage{graphicx}      
\usepackage{setspace}      
\usepackage{epstopdf}      
\usepackage{verbatim}      
\title[A Comparative Study of AGN Feedback Algorithms]{A Comparative Study of AGN Feedback Algorithms}
\author[J. Wurster and R.J. Thacker]{J. Wurster\thanks{E-mail: jwurster@ap.smu.ca} and R. J. Thacker\\
Department of Astronomy and Physics, St Mary's University, Halifax B3H 3C3, Canada\\}
\begin{document}

\date{Accepted 2013 February 21. Received 2013 February 20; in original form 2013 January 14}

\pagerange{\pageref{firstpage}--\pageref{lastpage}} \pubyear{2013}

\maketitle

\label{firstpage}

\begin{abstract}
Modelling AGN feedback in numerical simulations is both technically and theoretically challenging, with numerous approaches having been published in the literature.  We present a study of five distinct approaches to modelling AGN feedback within gravitohydrodynamic simulations of major mergers of Milky Way-sized galaxies.  To constrain differences to only be between AGN feedback models, all simulations start from the same initial conditions and use the same star formation algorithm.  Most AGN feedback algorithms have five key aspects: the black hole accretion rate, energy feedback rate and method, particle accretion algorithm, black hole advection algorithm and black hole merger algorithm.  All models follow different accretion histories, and in some cases, accretion rates differ by up to three orders of magnitude at any given time.  We consider models with either thermal or kinetic feedback, with the associated energy deposited locally around the black hole.  Each feedback algorithm modifies the region around the black hole to different extents, yielding gas densities and temperatures within $r \sim 200$ pc that differ by up to six orders of magnitude at any given time.  The particle accretion algorithms usually maintain good agreement between the total mass accreted by $\dot{M} \mathrm{d}t$ and the total mass of gas particles removed from the simulation, although not all algorithms guarantee this to be true.  The black hole advection algorithms dampen inappropriate dragging of the black holes by two-body interactions.  Advecting the black hole a limited distance based upon local mass distributions has many desirably properties, such as avoiding large artificial jumps and allowing the possibility of the black hole remaining in a gas void.  Lastly, two black holes instantly merge when given criteria are met, and we find a range of merger times for different criteria.  This is important since the AGN feedback rate changes across the merger in a way that is dependent on the specific accretion algorithm used.  Using the $M_\text{BH}$--$\sigma$ relation as a diagnostic of the remnants yields three models that lie within the one-sigma scatter of the observed relation and two that fall below the expected relation.  The wide variation in accretion behaviours of the models reinforces the fact that there remains much to be learnt about the evolution of galactic nuclei.

\end{abstract}

\section{Introduction}
\label{intro}

In the hierarchical model of galaxy formation, the largest galaxies are formed last.  Naively, we would expect the highest star formation rates (SFRs) and the activity from Active Galactic Nuclei (AGNs) to occur in these most massive galaxies.  However, observational evidence contradicts this, showing that in massive galaxies, the peak SFRs and peak AGN activity occurred at redshifts 1--2 (e.g. \citealp{SWKJH96}; \citealp{MFSGSF96}), and not today.  This reduction in activity from $z \sim 2$ to today was termed `downsizing' by \citet{CSHC96}.  One favoured explanation of downsizing is that during mergers, gas from the merger fuels both star formation and AGN activity (e.g. \citealp{Setal88}; \citealp{SSB05}); the feedback from the increased AGN activity then blows away all the gas, leading to a red and dead galaxy (e.g. \citealp{SDH05}).

The observational motivation for this picture is the evidence that a supermassive black hole resides at the centre of all galaxies with stellar spheroids (e.g. \citealp{KR95}; \citealp{FM00}), and that they did not evolve independently of one another.  The two strongest correlations are the relationship between the black hole mass and the stellar velocity dispersion ($M_\text{BH}$--$\sigma$; e.g. \citealp{SR98}; \citealp{FM00}; \citealp{Getal00}; \citealp{Tetal02}; \citealp{K03}; \citealp{Getal09}), and the black hole mass and the bulge mass ($M_\text{BH}$--$M_\text{b}$; \citealp{Metal98}; \citealp{MD02}; \citealp{MH03}).  A likely implication of these relations is that the AGN feedback is self-regulated:  Outflows from the black hole following a strong accretion event interact with the surrounding gas, inhibiting further accretion events, and hence limiting black hole growth (e.g. \citealp{SR98}; \citealp{F99a}; \citealp{SO04}).  Large X-ray cavities in the hot gas halo around AGN (e.g. \citealp{BVFEN93}; \citealp{Metal00}) suggest that some outflow events can be quite powerful; these cavities are likely formed by jets from the AGN (e.g. \citealp{MN07}).  Although there is significant evidence supporting the AGN scenario for explaining downsizing, it is not yet understood precisely how feedback energy couples to the surrounding gas; possible mechanisms are line radiation pressure (e.g. \citealp{CAK75}; \citealp{PSK00}), radiation pressure on dust grains (e.g. \citealp{MQT05}), Compton heating of infalling gas (e.g. \citealp{CO01}) and photo-ionisation pressure (\citealp{BM74}; \citealp{COS78}).

AGN feedback has been implemented in many numerical simulations (e.g. \citealt{KG05}; \citealt{DSH05}; \citealt{SDH05}; \citealt{TSC06}; \citealt{SSDH07}; \citealt{ONB08}; \citealp{KPN09}; \citealt{BS09}; \citealt{DQM11}; \citealt{DDST12}), but both theoretical understanding and numerical implementations are fraught with difficulties.  First, the spatial scales involved in studying AGN and the related feedback span many orders of magnitude, and information on all of these scales is needed simultaneously to fully and properly understand AGN and their feedback.  On the smallest scale is the Schwarzschild radius,
\begin{equation}
\label{Rschw}
r_\text{S} = \frac{2GM_\text{BH}}{c^2},
\end{equation}
where $M_\text{BH}$ is the mass of the black hole, $G$ is Newton's gravitational constant and $c$ is the speed of light; typical values range $r_\text{S} \approx 3\times 10^{6-8}$ km.  On the next larger scale is the Bondi radius \citep{B52}, 
\begin{equation}
\label{RBondi}
r_\text{Bondi} = \frac{GM_\text{BH}}{c_\infty^2},
\end{equation}
where $c_\infty$ is the sound speed of gas at infinity.  The value of the Bondi radius is dependent on the black hole's environment and can range from a few parsecs to tens of parsecs (e.g. \citealp{KPN09}; \citealt{SDH05}).  This radius, also known as the capture radius, divides a gas flow in to two regimes \citep{FKR02}.  Consider a spherically symmetric gas cloud centred on a black hole, where the gas is initially at rest at infinity.  The only forces acting on the gas are the gravitational force from the central black hole and the pressure forces within the gas (assuming we neglect the self-gravity of the gas).  Beyond the Bondi radius, the gas is comparatively uninfluenced by the black hole and flows subsonically.  Within $r_\text{Bondi}$, the gas density begins to increase, and the gas flow inward eventually reaches a sonic point.  At the sonic point, the gas plunges at a free-fall rate in to the black hole \citep{HPNK12}.  Thus, the Bondi radius can be viewed as the black hole's gravitational radius of influence.

The last spatial scale of interest is that of an entire galaxy (or a galaxy cluster), which can span dozens of kiloparsecs (or a few megaparsecs).  When considering all of these scales, comparing the size of a black hole to its massive host galaxy is similar to comparing a coin to the Earth \citep{F12}.

To complement the range of spatial scales, studying AGN feedback also requires a large range of temporal scales.  At short intervals, observations show the luminosity of the central engines of AGN varies on time-scales ranging from days to years (e.g. \citealp{WM00} and references therein; \citealp{SKKLPM11}); moreover, there are short term differences in variability amongst the different classes of AGN, making the variable luminosity challenging to understand and constrain.  Large scale observations have detected large ($\sim$10 kpc radius) X-ray cavities in the gas around the AGN.  To inflate these cavities, an outburst of $10^{58}$--$10^{61}$ ergs of energy would be required (\citealp{BRMWN04}; \citealp{MNetal05}) every few $\sim$$10^8$ yr, or a time-averaged output of $\sim$$10^{43}$--$10^{45}$ ergs s$^{-1}$ \citep{CSFB02}.  Thus, the next inherent difficulty in modelling AGN becomes obvious: AGN luminosity varies on time-scales as short as days, yet they are expected to produce major outbursts every few $\sim$$10^8$ yr.

Numerically, we can draw a few parallels between AGN feedback and stellar feedback.  Both have been added to numerical simulations to improve realism, and the result was better agreement with observations.  However, star formation and stellar feedback is a conceptually simpler problem than AGN feedback.  In numerical simulations, if a packet of gas meets a given set of criteria, then stars are formed and feedback energy is returned; the star formation and feedback parameters can be reasonably well constrained using current observations, where there are detailed observations of outflows at many different stages of stellar evolution, such as T-Tauri stars (e.g. \citealt{CESS90}; \citealt{HEG95}), Wolf-Rayet stars (e.g. \citealt{C07}), mass loss and stellar winds from evolved stars (e.g. \citealt{W00}) and supernovae (e.g. \citealt{R08}).  AGN feedback is much harder to constrain since there are fewer candidates for detailed observation due to the peak of AGN activity being so far in the past.  The nearest black hole candidate to study, Sagittarius A*, is currently in a quiescent phase; even during a recent flare to $4 \times 10^{39}$ erg s$^{-1}$ \citep{NRTK11}, its luminosity remained below the typical AGN luminosity range of $10^{40}$--$10^{47}$ erg s$^{-1}$ \citep{F99}.  In addition to uncertainty in modelling AGN feedback, tremendous amounts of energy can be numerically returned to the region around the back hole, creating steep gradients and presenting a challenge to numerical integration.  Unlike numerical stellar feedback where there is difficulty in preventing unrealistically rapid cooling after the feedback event (e.g. \citealp{GI97}; \citealp{MYTN97}; \citealp{TC00}; \citealp{SH03}), continual AGN feedback may create shocks or unrealistically disrupt the system.

When AGN feedback is included in numerical simulations, the approaches can vary widely, in both physics and numerical implementation.  Also, the physics is often modelled using different numerical codes with different star formation algorithms and different initial conditions.  Even starting from the same initial conditions, mass and spatial resolution, the different hydrodynamical approaches, gas cooling, star formation and stellar feedback algorithms of different numerical codes produce different results, as shown in the comparison by \citet{Setal12}; this comparison did not include AGN feedback.  Currently, a direct numerical comparison of the AGN feedback mechanisms does not appear to have been published nor can it be compiled from the assembled literature.  Here, we run four of the algorithms found in the literature using the numerical code {\sc Hydra}.  We also present a fifth simulation that is new to this study; this model is designed to take advantage of the `best' features of the other models.  All of our simulations start from the same initial conditions and use the same star formation algorithm; this approach is conservative, but well constrained.  We explicitly state that the goal of this is to highlight the different behaviours of the different algorithms and not to critique the various approaches.  

By using a pseudo-multiphase star formation model for all simulations, we unavoidably introduce some compromises in the AGN feedback models that rely upon a multiphase gas description; to compensate for this, we have introduced additional variables to represent hot and cold fractions of the gas.  Although this star formation model is not precisely equivalent to those implemented in other simulations of AGN feedback, we consider using it a necessary compromise to ensure that the variations in our results are only from the AGN feedback algorithm.  

The layout of this paper is as follows.  In Section \ref{sims} we discuss our simulations, focusing on how the initial conditions are constructed and the AGN feedback algorithms.  In Section \ref{results}, we will discuss our results, focusing on the impact of the different AGN feedback algorithms.  In Section \ref{fstates}, we discuss the final state of each simulation, and we conclude with a review in Section \ref{conclusion}.

\section{Numerical Simulations}
\label{sims}
To perform our simulations, we use the parallel version of {\sc Hydra} (\citealp{CTP95}; \citealp{TC06}), which uses an Adaptive Particle-Particle, Particle-Mesh algorithm \citep{C91} to calculate gravitational forces and the standard smooth particle hydrodynamics method (SPH; \citealp{GM77}; \citealp{L77}) to calculate gas forces.  It includes a star formation algorithm (see section \ref{simsSF}), and has been modified to include black holes and AGN feedback (see section \ref{simsBAM}).

\subsection{Galaxy models}
\label{simsGM}
To construct our model galaxy, we first use the GalactICs package (\citealp{KD95}; \citealp{WD05}; \citealp{WPD08}) to create a Milky Way-sized galaxy that consists of a stellar bulge, stellar disc, and a dark matter halo; this is done through an iterative process to produce a self consistent system.  The free parameters are chosen such that the component masses are similar to the component masses in \citet{SDH05}, and are given in Table \ref{galprops}.
\begin{center}
\begin{table}
{\small
\hfill{}
\begin{tabular}{l l l}
\hline
Component             & Parameter    	    & Value     	   \\
\hline
\hline
Bulge                 & $\sigma_\text{b}$   & 292   km s$^{-1}$    \\
                      & $R_\text{e}$        &   0.7 1kpc 	   \\
                      & $n$     	    &   1.1     	   \\
\hline
Disc                  & $R_\text{d}$        &   2.46 kpc 	   \\
                      & $z_\text{d}$        &   0.49 kpc 	   \\
                      & $R_\text{trunc}$    &  30    kpc    	   \\
                      & $z_\text{trunc}$    &   1    kpc    	   \\
                      & ${\sigma_\text{R}}_0$ & 119    km s$^{-1}$   \\
\hline
Dark Matter Halo      & $a_\text{h}$        &  13.6  kpc  	   \\
                      & $r_\text{h}$        & 275    kpc    	   \\
                      & $\delta r_\text{h}$ &  25    kpc    	   \\
                      & $\sigma_\text{h}$   & 330    km s$^{-1}$   \\
                      & $\gamma$     	    &   0.81    	   \\
\hline
metallicity           & $Z$                 & 0.05 Z$_{\astrosun}$ \\
mean molecular weight & $\mu$               & 0.6       	   \\
\hline
\end{tabular}}
\hfill{}
\caption{The chosen parameters for our model galaxies.  All parameters are defined in Section \ref{simsGM} of the text.}
\label{galprops} 
\end{table}
\end{center}

The stellar bulge density profile is given by
\begin{equation}
\label{GICbulge}
\tilde{\rho}_\text{b}(r) = \rho_\text{b} \left( \frac{r}{R_\text{e}} \right)^{-p} \text{e}^{-b(r/R_\text{e})^{1/n}},
\end{equation}
which yields a S\'ersic law for the projected density profile if $p = 1 - 0.6097/n + 0.05563/n^2$, where $n$ is a free parameter.  The constant $\rho_\text{b}$ is defined using $\sigma_\text{b} \equiv \left\{ 4\pi n b^{n(p-2)}\Gamma \left[n(2-p)\right]R_\text{e}^2 \rho_\text{b}\right\}^{1/2}$, where $\sigma_\text{b}^2$ is the depth of the gravitational potential associated with the bulge and $R_\text{e}$ is the radial scale parameter, which is a free parameter; the variable $b$ is adjusted such that $R_\text{e}$ encloses half the total projected light or mass.  

The stellar disc has a truncated density profile that falls off approximately exponentially in $R$ and follows sech$^2$ in $z$; the disc has radial and vertical scale heights $R_\text{d}$ and $z_\text{d}$ and truncation distances $R_\text{trunc}$ and $z_\text{trunc}$.  The radial velocity dispersion is given by $\sigma_\text{R}^2(R) = {\sigma^2_\text{R}}_0 \text{e}^{-R/R_\sigma}$, where ${\sigma_\text{R}}_0$ is the central velocity dispersion and $R_\sigma = R_\text{d}$ for simplicity.  

The dark matter halo profile follows
\begin{equation}
\label{GIChalo}
\tilde{\rho}_\text{h} = \frac{2^{2-\gamma}\sigma^2_\text{h}}{4\pi a^2_\text{h}} \frac{1}{(r/a_\text{h})^\gamma (1+r/a_\text{h})^{3-\gamma}} C(r;r_\text{h},\delta r_\text{h}),
\end{equation}
where $a_\text{h}$ is the halo scale length, $r_\text{h}$ is the cutoff radius, $\gamma$ is the central cusp strength and $\sigma_\text{h}$ is a (line of sight) velocity scale that sets the mass of the halo.  The truncation function, 
$C(r;r_\text{h},\delta r_\text{h}) = \frac{1}{2}\text{erfc}\left( \frac{r-r_\text{h}}{\sqrt{2}\delta r_\text{h}} \right)$, smoothly goes from one to zero at $r = r_\text{h}$ over width $\delta r_\text{h}$.  

We then modify this initial galaxy in three ways.  First, we convert ten per cent of the total stellar mass into gas to create a gas disc; the gas disc is the same as the stellar disc except that it has been reflected in the $x = y$ plane to avoid coincidence with the star particles.  Although the gas scale height is initially larger than physically motivated, cooling allows the gas to collapse into a thin disc within a few 10 Myr; this produces a short transient evolution of the gas accompanied by a brief increase in the SFR.  At resolutions higher than presented here, this vertical collapse produces a strong ring-shaped shock which propagates outwards.  One solution is to relax the gas disc in a fixed potential and then implement the relaxed disc in the initial conditions; another solution is to reduce the initial scale height of the gas disc.  Studies in \citet{WT12} show that at our fiducial resolution the pressure wave dissipates comparatively quickly, and is largely insignificant at our lowest resolutions.  Given also that we are interested in the relative differences between the feedback algorithms during the merger, which is clearly a much more dynamic event, we consider the above gas disc construction acceptable.

Second, we add a hot gas halo (hgh), which is chosen to follow the observationally motivated $\beta$-profile (e.g. \citealp{CF76}):
\begin{equation}
\label{betaprofile}
\rho_\text{hgh}(r) = \rho_0\left[1 + \left(\frac{r}{r_\text{c}}\right)^2\right]^{-\frac{3}{2}\beta},
\end{equation}
where $\rho_0$ is the central density, $r_\text{c}$ is the core radius, and $\beta$ is the outer slope parameter; we choose $r_\text{c} = 1.75$ kpc and $\beta = 2/3$ as done in \cite{MMSNC11}.  We set $\rho_0$ by choosing the mass of the hot gas halo within 40 kpc to be equal to two per cent of the total disc mass \citep{Retal09}; to conserve total halo mass generated by GalactICs, we reduce the mass of the dark matter halo by the total mass of the hot gas halo.  By assuming isotropy and hydrostatic equilibrium, the temperature profile of the hot gas halo is given by \citep{KMWSM07}
\begin{equation}
\label{hghtemp}
T_\text{hgh}(r) = \frac{\mu m_\text{p}}{k_\text{B}} \frac{1}{\rho_\text{hgh}(r)} \int_r^\infty  \rho_\text{hgh}(r) \frac{GM(r)}{r^2}  \mathrm{d} r, 
\end{equation}
where $\mu$ is the mean molecular weight, $m_\text{p}$ is the proton mass, $k_\text{B}$ is the Boltzmann constants, and $M(r)$ is the total mass interior to $r$.  The hot gas halo is given an initial angular momentum which scales with the circular velocity, $j_\text{hgh}(R) \propto R v_\text{circ}(R)$, where $R$ is the distance from the spin axis of the galaxy \citep{MMSNC11}.  

The final modification to the galaxy is the addition of a black hole (sink) particle, which is placed at the centre of mass and given an internal mass of 10$^5$ M$_{\astrosun}$.  While this initial mass is more than ten times lower than expected from estimates of the mass of the Milky Way’s central black hole ($M_\text{BH} \sim (4.36 \pm 0.42)\times 10^6$ M$_{\astrosun}$; \citealp{GETetal09}), it does match the initial black hole masses in \citet{SDH05} and \citet{DQM11}. Moreover, an initial mass lower than anticipated from the $M_\text{BH}$--$\sigma$ relationship would be expected to grow quite quickly in a few Salpeter times because feedback at low masses is comparatively weak. At every step, we calculate a ‘smoothing length’ for the black hole in order to calculate the gas properties around it, but the sink particle itself only experiences gravitational forces. 

Each galaxy has a total mass of $9.60\times 10^{11}$ M$_{\astrosun}$, 1 287 743 (168 351) particles for the fiducial (low) resolution simulations, and a Plummer softening length of $\epsilon_\text{Plummer} = 120$ pc ($\epsilon_\text{Plummer} = 300$ pc).  The Plummer softening length is related to the S2 gravitational softening length, $\epsilon_\text{S2}$, by $\epsilon_\text{Plummer} = \epsilon_\text{S2}/2.34$.  See Table \ref{breakdown} for a breakdown of each galaxy.  Lastly, we create a second, identical galaxy, and separate them by 70 kpc; we then place them on parabolic trajectories around one another, in close agreement with the trajectories from \citet{SDH05}.
\begin{center}
\begin{table*}
\begin{minipage}{\textwidth}
{\small
\hfill{}
\begin{tabular}{l|r r r r r r}
    \hline
  &                           & \multicolumn{2}{c}{Fiducial resolution} &&  \multicolumn{2}{c}{Low resolution} \\
\cline{3-4}
\cline{6-7}
 & \multicolumn{1}{c}{Total mass}                 & \multicolumn{1}{c}{Particle mass}           & \multicolumn{1}{c}{Number of}  && \multicolumn{1}{c}{Particle mass}           &  \multicolumn{1}{c}{Number of}\\
 & \multicolumn{1}{c}{(10$^{10}$ M$_{\astrosun}$)}& \multicolumn{1}{c}{(10$^5$ M$_{\astrosun}$)}& \multicolumn{1}{c}{particles}  && \multicolumn{1}{c}{(10$^5$ M$_{\astrosun}$)}&  \multicolumn{1}{c}{particles}\\
\hline
\hline
Dark matter halo & 89.92    & 11.75 & 765 000 && 89.92 & 100 000 \\
Hot gas halo     &  0.60    &  0.36 & 165 343 &&  2.77 &  21 619 \\
Stellar bulge    &  1.34    &  2.37 &  56 649 && 18.10 &   7 407 \\
Stellar disc     &  3.56    &  2.37 & 150 375 && 18.10 &  19 662 \\
Gas disc         &  0.54    &  0.36 & 150 375 &&  2.77 &  19 662 \\
Black hole$^\text{a}$
	         &10$^{-5}$ &  1.00 &       1 &&  1.00 &       1 \\
\hline
\end{tabular}}
\hfill{}
\caption{Component breakdown for each galaxy. \hspace{8in} \newline$^\text{a}$Particle (dynamical) mass is 10$^9$ M$_{\astrosun}$ for Model DQM; internal mass of the black hole remains at 10$^5$ M$_{\astrosun}$.}
\label{breakdown} 
\end{minipage}
\end{table*}
\end{center}

\subsection{Verification against other codes}
\label{simsV}
Since many of the models in the literature are run using {\sc Gadget2} (\citealp{SYW01}; \citealp{SH02}), we ran the same adiabatic merger simulation with both {\sc Hydra} and the publicly available version of {\sc Gadget2} to test the differences.  Neither simulation included stellar or AGN feedback, thus we were essentially comparing the  gravity and SPH solvers.  The galaxies in each simulation followed the same trajectory, and synchronisation -- as measured by the time to reach second periapsis -- was within 0.2 per cent. Thus we can be confident that our results will be comparable with those currently found in the literature. 

\subsection{Star formation}
\label{simsSF}
The star formation algorithm implemented in {\sc Hydra} is described and tested in \cite{TC00}; we will provide a brief summary here. This algorithm follows an approach to star formation that has now been studied extensively in the literature, where star formation is allowed to proceed in regions where:
\begin{enumerate}
  \item The gas exceeds the density limit of $n_\text{H} \sim 0.01$ cm$^{-3}$,
  \item The flow is convergent, ${\bf \nabla} \cdot {\bf v} < 0$,
  \item The gas temperature is less than $3\times 10^4$ K,
  \item The gas is partially self-gravitating: $\rho_\text{g} > 0.4\rho_\text{DM}$.
\end{enumerate}
When a gas particle meets the above criterion, the Schmidt Law (e.g. \citealp{k92}; \citealp{K98}) is used to determine the amount of gas that is `converted' in to stars.  When the cumulative converted mass of a gas particle reaches half of the particle's original mass, $m_\text{g}$, a star particle is spawned with mass $m_\text{g}/2$ and the gas particle's mass is reduced to $m_\text{g}/2$; when 80 per cent of the remaining gas mass is converted into stars, a second star particle is spawned with mass $m_\text{g}/2$ and the gas particle is removed from the simulation.  For computational efficiency, {\sc Hydra} uses a Lagrangian form of the Schmidt Law, based upon the local density of the SPH particle,
\begin{equation}
\frac{\mathrm{d}M_*}{\mathrm{d}t} = C_\text{sfr}\rho_\text{g}^{1/2}M_\text{g},
\end{equation}
where $C_\text{sfr}$ is the star formation rate normalisation, $\rho_\text{g}$ is the SPH density of the gas particle, and $M_\text{g}$ is the mass of the gas in the particle that has not been converted in to stars.  All other calculations in the code (e.g. density, smoothing length, etc...) assume that the entire particle is a gas particle of mass $m_\text{g}$ or $m_\text{g}/2$. 

Whenever a star particle is created, feedback energy is immediately returned to the surrounding environment.  Although our fiducial resolution models takes $\mathcal{O}(10^2)$ steps to evolve through the lifetime of an 8 M$_{\astrosun}$ star, the lack of delay between star particle formation and feedback energy release is not a significant issue because at the resolution considered here we are still averaging over a number of giant molecular clouds per particle and the evolution is representative at best.  For every 100 M$_{\astrosun}$ of stars formed, there is one supernova event, which contributes 10$^{51}$ ergs to the interstellar medium \citep{SGV99}, and feeds back $5 \times 10^{15} e^*$ ergs g$^{-1}$ of gas converted in stars, where $e^*$ is a dimensionless parameter; we set $e^* = 0.4$ to match \cite{NW93}.  

As first showed by \cite{k92}, feedback energy returned to the interstellar medium (ISM) is radiated away quickly in high density regions.  A reduced density is used in the cooling algorithm to prevent this immediate loss, thus allowing the feedback to influence the surrounding environment; the reduced density decays back to its local SPH value with a half-life of $t_{1/2} = 5$ Myr.  The parameters $e^*$ and $t_{1/2}$ are chosen to match those set in \citet{TC00} to reproduce the Milky Way's star formation rate in a simulation of an isolated Milky Way-like galaxy. 

This star formation algorithm is conceptually similar to the methods used in \citet{BKGF04}, \citet{SSKWGQ06} and \citet{Getal07}, however, it varies from the sub-resolution multiphase models found in \citet{SH03}, \citet{SDH05}, \citet{BS09} and \citet{HCYL09}.  In multiphase models, the subgrid physics is derived from a model of cold clouds (where stars form) embedded in a hot, pressure-confining phase.  Above a given density threshold, the gas is thermally unstable to the onset of this two-phase medium.  The mass fraction in each phase is determined by star formation and feedback, evaporation of the cold clouds through thermal conduction, and the growth of clouds through radiative cooling.  The star formation rate is then calculated based upon a prescribed law and an equation of state.  A noteworthy difference between this method and our method is that here, gas can freely flow between phases, whereas in {\sc Hydra}, once gas is `converted' to stars, it is carried forward and not allowed to cool until the specified cooling period is reached.

\subsection{Black hole and AGN feedback algorithms}
\label{simsBAM}
AGN feedback algorithms essentially have five key components: 
\begin{enumerate}
  \item The accretion rate on to the black hole,
  \item The SPH particle accretion algorithm,
  \item The energy feedback algorithm,
  \item The black hole advection algorithm, and
  \item The black hole merger algorithm.
\end{enumerate}
Each component will briefly be discussed below, then we will discuss the models in section \ref{Nmodels}.

\subsubsection{Accretion rates}
A commonly used accretion rate is the Bondi accretion rate 
\citep{B52},  
\begin{equation}
\label{bondi}
\dot{M}_\text{Bondi} = \frac{2 \pi G^2 M_\text{BH}^2\rho_\infty}{(c_\infty^2+v^2)^{3/2}},
\end{equation}
where $\rho_\infty$ and $c_\infty$ are the gas density and sound speed at infinity, $v$ is the relative velocity between the gas at infinity and the black hole, and $M_\text{BH}$ is the mass of the black hole.  In hydrostatic equilibrium, the maximum physical accretion rate is the Eddington accretion rate, 
\begin{equation}
\label{mEdd}
\dot{M}_\text{Edd} \equiv \frac{4 \pi G M_\text{BH} m_\text{p}}{\epsilon_\text{r} \sigma_\text{T} c},
\end{equation}
where $m_\text{p}$ is the proton mass, $\sigma_\text{T}$ is the Thomson cross section, and $\epsilon_\text{r}$ is the radiative efficiency (i.e. the mass-to-energy conversion efficiency); we set $\epsilon_\text{r} = 0.1$ \citep{SS73}.  Typically, numerical simulations limit their accretion rate to the Eddington accretion rate, and we follow this convention.  If the assumption of spherical symmetry is ignored, super-Eddington accretion rates can be achieved, which are associated with collimated outflows.  In small scale outflow simulations by \citet{KPN09}, they find steady-state results where super-Eddington accretion occurs in the equatorial plane. 

\subsubsection{Black hole mass growth and particle accretion}
In all cases, the `internal' and `dynamical' masses of the black hole is tracked.  The internal mass, $M_\text{BH}$, is the mass of the black hole, which is increased by $\dot{M}_\text{BH} \mathrm{d}t$ at every iteration; this mass is used in all calculations concerning AGN feedback.  The dynamical mass, $m_\text{BH}$, is the mass of the sink particle, which is increased by the mass of a gas particle whenever one is accreted.  The particle accretion algorithm should ideally maintain $M_\text{BH} \sim m_\text{BH}$ and directly address the loss of gas near the black hole due to accretion.  In Model DQM (see \S \ref{Mdqm}), the dynamical mass is fixed for all time (except during a black hole merger), so the included particle accretion algorithm only simulates the loss of gas.  When a particle is accreted on to the black hole, its mass and momentum are added to the black hole particle (except for Model DQM), and the gas particle is removed from all further computations; this accretion does not affect the internal mass of the black hole.  There are three categories of particle accretion algorithms:
\begin{enumerate}
\item \emph{Stochastic-Unconditional}: At all times, nearby particles are tested to see if they will stochastically accrete.
\item \emph{Stochastic-Conditional}:   Nearby particles are tested to see if they stochastically accrete if given criteria of the $m_\text{BH}$--$M_\text{BH}$ relationship are satisfied. 
\item \emph{Continual-Conditional}:    Nearby particles are continually accreted while given criteria of the $m_\text{BH}$--$M_\text{BH}$ relationship are satisfied.
\end{enumerate}

\subsubsection{Feedback}
In all simulation, a portion of the accreted mass is returned as feedback energy, $E = \epsilon \dot{M}_\text{BH}c^2 \mathrm{d}t$, where $\epsilon$ is a dimensionless efficiency parameter.  This energy is returned to the nearby particles (except Model ONB; see \S \ref{Monb}) either by increasing their internal energy (thermal feedback models), or by increasing their momentum via $p = E/c$ (kinetic feedback models).

\subsubsection{Black hole advection}
Properly tracking the black hole particle is critical since accretion rates (hence feedback rates) depend on the local gas properties around the black hole.  When the black hole mass is similar to a star or gas particle mass, it can be easily and inappropriately dragged around by two-body forces leading to the inaccurate calculation of gas properties.  Avoiding such behaviour is clearly desirable.  In all of the simulations, a black hole advection algorithm is implemented to minimise the inappropriate movement; except in Model DQM (see \S \ref{Mdqm}), the black hole's position is artificially updated after the completion of the gravitational solver algorithm but before the calculation of the accretion rate.  

\subsubsection{Black hole mergers}
The process by which black holes merge is still a matter of active research (e.g. \citealp{ELCM04}; \citealp{BPBMS09}; \citealp{KJM11}; \citealp{BBHHLS12}).  Given that our simulations result in a combined halo and stellar system producing drag, it is reasonable to expect that the black holes would merge.  Since any merger occurs on sub-resolution scales, models include a merger prescription to instantly merge the black holes when given criteria are met.  When two black holes merge, one sink particle is removed and the remaining sink particle has the combined mass (both internal and dynamical) and momentum, and is repositioned to the centre of mass of the two progenitors.

\subsubsection{Black hole's local environment}
Around every black hole we define a radius of influence, $r_\text{inf}$.  All gas particles within $r_\text{inf}$ contribute to the accretion properties at the black hole, are eligible to receive feedback energy (except for Model ONB), and are eligible to be accreted on to the black hole particle.  Gas particles outside of $r_\text{inf}$ have no explicit impact on the black hole or the AGN feedback algorithms.  We set $r_\text{inf} = \max(2h_\text{BH},2h_\text{min})$, where a sphere with radius $2h_\text{BH}$ around the black hole particle includes 60 gas particles, and $h_\text{min}$ is the smallest resolved smoothing length in the SPH solver.

\subsection{The Models}
\label{Nmodels}
In sections \ref{Msdh} to \ref{Mwt}, we describe the particular AGN feedback algorithms of our five primary models; a summary of each model can be found in Table \ref{models}.  We ran four additional models, each of which is a slight variant of a primary model; these are described in section \ref{Madd}.  Every model was run at both fiducial and low resolutions.  For nomenclature, we name these models after the initials of the authors of the paper it originally appeared in.  The variant models have the same name as their parent model, followed by a lower case character to signify the difference.  A model name followed by a subscript `l' explicitly refers to the low resolution version of the model; no subscript will refer to either the fiducial resolution version or to both versions, depending on context. 

\begin{center}
\begin{table*}
\begin{minipage}{\textwidth}
{\tiny
\hfill{}
\begin{tabular}{l|l|l| l| l| l}
    \hline
 & Model SDH & Model BS & Model ONB & Model DQM & Model WT \\
\hline
\hline
Accretion Rate     & $\dot{M}_\text{B} (\alpha = 100)$& $\dot{M}_\text{B} (\alpha \equiv \alpha(n_\text{H}))$ 
							                           & $\dot{M}_\text{drag}$ 		& $\dot{M}_\text{visc}$		 & $\dot{M}_\text{B} (\alpha = 100)$ \\
																					\\
Energy Feedback    & $\dot{E} = 0.005\dot{M}_{BH} c^2$ 
						 & $\dot{E} = 0.015\dot{M}_\text{BH} c^2$ 
 										   & $\dot{E} = 0.1 L_\text{jet}$ 	& $\dot{p} = 10 L/c$ 	 	  & $\dot{E} = 0.005 \dot{M}_\text{BH} c^2$      	\\
FB Distribution    & kernel-weighted              & random packets of $E_\text{crit}$    & to 40 low-$\rho$ particles    & isotropic			  & isotropic 		    				\\
																							    			\\
Particle Accretion & stochastic-unconditional    & stochastic-conditional	   & continual-conditional 		& continual-conditional		  & continual-conditional				\\
																							    			\\
BH Advection       & gas particle with 		 & gas particle with 		   & $\Delta l_\text{ONB}$ 		& tracer mass		    	  & $\Delta l_\text{WT}$   	    			\\
                   & $v_\text{rel} < 0.25c_\text{s}$  	 & with $v_\text{rel} < 0.25c_\text{s}$        & along stellar gradients    & 			  & towards centre of mass  				\\
                   & and smallest $U$ 	 	 & and smallest $U$		   &                            	& 			    	  &             	    				\\
																			  	            					\\
BH Merger          & $d < h_\text{BH}$  	 & $d < h_\text{BH}$  		   & $d < \epsilon_{S2}$        	& $       d < h_\text{BH}$ 	  & $d < h_\text{BH}$    			        \\
                   & \& $v_\text{rel} < c_\text{s}$      & \& $v_\text{rel} < v_{circ}$	   & \& gravitationally bound   &                                 & \& $v_\text{rel} < c_\text{s}$ 		     	\\
																					  		    			\\
\hline
\end{tabular}}
\hfill{}
\caption{A summary of our five primary models describing each key algorithm.  More detail and definitions of the variables are given in subsections \ref{Msdh} to \ref{Mwt}.  Numerically, $h_\text{BH} = \max(h_\text{BH},h_\text{min})$, where a sphere with radius $2h_\text{BH}$ around the black hole particle includes 60 gas particles, and $h_\text{min}$ is the smallest resolved smoothing length in the SPH solver.}
\label{models} 
\end{minipage}
\end{table*}
\end{center}

\subsubsection{Model SDH}
\label{Msdh}
This model is based upon that found in \cite{SDH05} (herein SDH05).  The accretion rate is given by a modified Bondi accretion rate,
\begin{equation}
\label{mBondi}
\dot{M}_\text{B} = \frac{4\pi\alpha G^2 M^2_\text{BH} \rho}{\left(c_\text{s}^2 + v_\text{rel}^2\right)^{3/2}},
\end{equation}
where $c_\text{s}$ and $\rho$ are the local sound speed and density of the gas, and $v_\text{rel}$ is the relative velocity of the black hole to the nearby gas.  The free parameter, $\alpha$, is included to relate the numerically calculated gas density and sound speed to what one would expect in reality.  \citet{BS09} argue that modest resolutions underestimate the density and overestimate the sound speed by orders of magnitude, thus justifying large values of $\alpha$.  As in SDH05, we set $\alpha = 100$.  Finally, the accretion rate is Eddington-limited, thus $\dot{M}_\text{BH} = \min\left( \dot{M}_\text{B}, \dot{M}_\text{Edd} \right)$.

A given fraction of the accreted mass is allowed to return to the surrounding environment as feedback energy.  The rate of return is 
\begin{equation}
\label{Efeed1}
\dot{E}_\text{feed} = \epsilon_\text{f} \epsilon_\text{r} \dot{M}_\text{BH} c^2, 
\end{equation}
where $\epsilon_\text{r} = 0.1$ is the radiative efficiency, and $\epsilon_\text{f} = 0.05$ is the fraction of energy that can couple with the gas.  This energy is returned kernel-weighted to the gas particles within $r_\text{inf}$.

To track the growth of the black hole particles, a stochastic-unconditional particle accretion algorithm is used.  At every iteration, a probability, $p_i$, is calculated for each particle, $i$, within $r_\text{inf}$.  This probability is then compared to a random number, $x_i$, and if $p_i > x_i$, then the particle is accreted.  In this algorithm, the probability is given by 
\begin{equation}
\label{prob1}
p_i = w_i \dot{M}_\text{BH} \rho^{-1} \mathrm{d}t,  
\end{equation}
where $w_i$ is the kernel weight of gas particle $i$ relative to the black hole, and $x_i$ is uniformly distributed on the the interval (0,1).

To minimize inappropriate motions of the black hole particles, at every iteration the black hole is relocated to the nearby gas particle with the minimum potential energy provided that $v_\text{rel} < 0.25c_\text{s}$.  If no gas particle meeting the velocity criteria exists within $r_\text{inf}$, then the black hole is not advected.  Once $m_\text{BH} > 10m_\text{g}$, the black hole advection is turned off and its movement is handled only by the gravitational solver. 

Lastly, two black holes merge when they come within each other's smoothing lengths and have a relative velocity less than the local sound speed; SDH05 argue that the local sound speed represents a simple measure of the velocity scale at which the black holes are able to merge.

\subsubsection{Model BS}
\label{Mbs}
This model is based upon that found in \cite{BS09} (herein BS09); this model has many similarities to Model SDH, but was originally implemented in a cosmological simulation.  We caution that implementing their model in a higher-resolution simulation could lead to unwanted behaviours as a number of the model parameters were chosen for their specific resolution.  This model uses the (modified) Bondi accretion rate given in \eqref{mBondi}.  If the resolution is sufficient enough, BS09 argue that the justifications for $\alpha = 100$ given in section \ref{Msdh} break down in low-density regions.  Thus, this model sets $\alpha$ to be a function of the local hydrogen density, $n_\text{H}$:
\begin{equation}
\label{alpha}
\alpha = \left\{ \begin{array}{c l} 1 		    & \text{if   } n_\text{H} < n_\text{H}^* \\
\left(\frac{n_\text{H}}{n_\text{H}^*}\right)^\beta  & \text{otherwise}
\end{array}\right., 
\end{equation}
where $n_\text{H}^*$ is the critical value required for the formation of a cold interstellar gas phase, and $\beta$ is a free parameter; as in BS09, we set $n_\text{H}^* = 0.1$ cm$^{-3}$ and $\beta = 1$.  Finally, the accretion rate is Eddington-limited, thus $\dot{M}_\text{BH} = \min\left( \dot{M}_\text{B}, \dot{M}_\text{Edd} \right)$.

The rate of feedback is also given by \eqref{Efeed1}, with $\epsilon_\text{r} = 0.1$ and $\epsilon_\text{f} = 0.15$, which is three times more efficient than Model SDH.  The feedback energy is allowed to accumulate until $E > E_\text{crit}$, at which point a random gas particle within $r_\text{inf}$ receives $E_\text{crit}$ energy; this is repeated until the accumulated energy drops below $E_\text{crit}$. The critical energy is defined as 
\begin{equation}
\label{Ecrit}
E_\text{crit} = \frac{m_\text{g} k_\text{B} \Delta T}{\left(\gamma - 1\right) \mu m_\text{H}},
\end{equation}
where $m_\text{g}$ is the (initial) mass of a gas particle and $\Delta T$ is the temperature increase a particle experiences with every feedback event.  Due to the higher resolution in our model ($m_\text{gas, \ fiducial} = 3.6\times 10^4$ M$_{\astrosun}$ compared to $8.64\times10^7$ M$_{\astrosun} \ h^{-1}$ in BS09; $\epsilon_\text{fiducial} = 120$ pc compared to 2 kpc $h^{-1}$ after $z = 2.91$ in BS09), we set $\Delta T = 5\times 10^6$ K as opposed to $\Delta T = 10^8$ K used in BS09.  Using the value in BS09 produces very large temperature gradients in the SPH solver that are difficult to integrate accurately.  

Gas particles are accreted by a stochastic-conditional particle accretion algorithm.  If $M_\text{BH} < m_\text{BH}$, then the probability of accretion is $p_i \equiv 0$, otherwise it is calculated using
\begin{equation}
\label{prob2}
p_i = w_i \left(M_\text{BH} - m_\text{BH}\right)\rho^{-1}.
\end{equation}
As in Model SDH, particle $i$ is accreted if $p_i > x_i$.

The black hole advection is the same as in Model SDH.  Lastly, two black holes merge when they come within each other's smoothing lengths and have a relative velocity less than the circular velocity at the radius of the most massive black hole's smoothing length.  BS09 explicitly state that this merging criteria differs from SDH05 since the feedback returned may temporarily increase the local sound speed, thus may not be a representative velocity scale.

\subsubsection{Model ONB}
\label{Monb}
This model is based upon the model found in \cite{ONB08} (herein ONB08), which was originally implemented in a simulation with cosmological initial conditions.  We also note that this model is distinctly different from our other models in that it is specifically designed to reproduce the radio mode of feedback.  In this model, it is assumed that radiation from stars (e.g. through starbursts or winds) interacts with the rotating, clumpy ISM.  This radiation irradiates one layer of gas at a time, extracting angular momentum from it.  This permits an inflow of gas towards the galactic centre, and ultimately on to the black hole itself (\citealp{UFM97}; \citealp{U01}; \citealp{KU02}).  Thus, an accretion rate of gas on to the black hole can be calculated by considering the stellar clouds in the region of star formation (RSF) near the black hole.  Using these assumptions, ONB08 calculate the drag accretion to be
\begin{equation}
\label{drag}
\dot{M}_\text{drag} = \epsilon_\text{drag} \frac{L_\text{RSF}}{c^2}\left(1-\text{e}^{-\tau_\text{RSF}}\right),
\end{equation}
where $\epsilon_\text{drag} = 1$ is the drag efficiency, $L_\text{RSF}$ is the total bolometric luminosity of all the stars in the RSF, and $\tau_\text{RSF}$ is the total optical depth of the RSF.  The total luminosity is calculated by summing the age-dependent bolometric luminosities, which are obtained from a lookup table generated by {\sc pegase2} \citep{FR97}.  Next, the total optical depth of a gas cloud is given by $\tau_\text{c} = \chi_\text{d} \rho_\text{c} r_\text{c} \simeq \chi_\text{d} m_\text{c}/r_\text{c}^2$ where $\rho_\text{c}$, $m_\text{c}$ and $r_\text{c}$ are the density, mass and radius of the cloud, respectively.  Assuming that all the clouds are identical and randomly distributed over the region of star formation, the total optical depth can be approximated by
\begin{equation}
\label{tau}
\tau_\text{RSF} = \frac{3\chi_\text{d}}{4\pi} \frac{M_\text{c}}{R_\text{RSF}^2},
\end{equation}
where $\chi_\text{d} = 50$ cm$^2$ g$^{-1}$ is the mass extinction coefficient, $M_\text{c}$ is the total mass of the clouds in the RSF, and $R_\text{RSF}$ is the radius of the RSF; since {\sc Hydra} does not explicitly track a multiphase gas, we set $M_\text{c}$ to be half of the total gas mass within the RSF.  We initialise $R_\text{RSF} = \max\left(R_\text{40},2h_\text{min}\right)$, where a sphere with radius $R_\text{40}$ centred on the black hole contains 40 gas particles.  Then, larger radii are searched to find the radius that maximises $\dot{M}_\text{drag}$.  Lastly, if the gas density within the sphere is less than $\rho_\text{thresh} = 5\times 10^{-25}$ g cm$^{-3}$, the accretion rate is set to zero.

In this model, it is explicitly assumed that the feedback heats the halo gas through the production of jets.  The jet mechanism used here is that of \cite{M01}, and generates power from the rotational energy of the accretion flow and from the black hole itself.  The accretion flow is divided into two regimes: standard thin discs (SD; optically thick, geometrically thin, radiatively efficient) and radiatively inefficient accretion flows (RAIF; optically thin, geometrically thick).  Using the parameters in ONB08, the respective accretion-dependent luminosities are
\begin{align}
\label{Efeed4}
L_\text{jet}^\text{SD  } & \approx 8.1 \times 10^{-5}\dot{M}_\text{BH} c^2 & \text{if   } \dot{m}  >  \dot{m}_\text{crit} \\
L_\text{jet}^\text{RAIF} & \approx 2.6 \times 10^{-1}\dot{M}_\text{BH} c^2 & \text{if   } \dot{m} \le \dot{m}_\text{crit},
\end{align}
where $\dot{m} \equiv \dot{M}_\text{BH}/\dot{M}_\text{Edd}$ and $\dot{m}_\text{crit} \approx \alpha^2$ is the critical accretion rate that sets the division between the SD and RAIF regimes (e.g. \citealp{NMQ98}).  As in ONB08, we set $\alpha_\text{SD} = \alpha_\text{RAIF} \equiv \alpha =0.1$, where $\alpha_\text{SD}$ and $\alpha_\text{RAIF}$ are the viscosity parameters for SD and RAIF's, respectively.  The feedback rate is then given by $\dot{E}_\text{feed} = \epsilon_\text{r} L_\text{jet}$, and the energy is distributed equally to the 40 nearest diffuse gas particles with $\rho < 0.1 \rho_\text{thresh}$. 

To accrete particles, this model uses a continual-conditional particle accretion algorithm; when the internal black hole mass exceeds its dynamical mass, nearby gas particles are accreted with the probability
\begin{equation}
\label{prob4}
p_i = \frac{M_\text{BH}-m_\text{BH}}{m_i N_\text{RSF}},
\end{equation}
where $N_\text{RSF}$ is the number of gas particles within the region of star formation, and $m_i$ is the mass of the $i$'th gas particle; particles are accreted until the dynamical mass exceeds the internal mass.

To track the black hole, at every iteration, the local stellar density fields are computed and the black hole is moved along the steepest gradient by an amount 
\begin{equation}
\label{deltal4}
\Delta l_\text{ONB} = \min(0.01\epsilon_\text{S2}, 0.03 \left| \boldsymbol{v} \right| \mathrm{d}t),  
\end{equation}
where $\epsilon_\text{S2}$ is the gravitational softening length, $\left| \boldsymbol{v} \right|$ is the velocity of the black hole, and $\mathrm{d}t$ is the time-step; these coefficients are the same as in ONB08 and were determined empirically.  

Lastly, two black holes are assumed to merge when they are within a softening length of one another and are gravitationally bound.

\subsubsection{Model DQM}
\label{Mdqm}
This model is based upon that found in \cite{DQM11} (herein DQM11), and models accretion through the transportation of angular momentum, which is based upon multi-scale SPH simulations by \cite{HQ10}.  The accretion rate is 
\begin{equation}
\label{Visc}
\dot{M}_\text{visc} = 3 \pi \delta \Sigma \frac{c_\text{s}^2}{\Omega}, 
\end{equation}
where $\delta$ is the dimensionless viscosity, $\Sigma$ is the mean gas surface density, and $\Omega = \sqrt{GM/r_\text{inf}^3}$ is the rotational angular velocity of the gas.  DQM11 treat $\delta$ as a free parameter, and it characterises both the efficiency of angular momentum transport and the fraction of gas that is being converted into stars versus being accreted on to the black hole; as in DQM11, we set $\delta = 0.05$.

The feedback is returned as momentum using
\begin{equation}
\label{Efeed3}
\dot{p} = \tau \frac{L}{c},
\end{equation}
where $\tau$ is the infrared optical depth, and $L = \min \left( \epsilon_\text{r} \dot{M}_\text{visc} c^2 , L_\text{Edd} \right)$ is the Eddington-limited luminosity.  This momentum outflow is used to approximate radiation pressure produced by absorption and scattering of the AGN's feedback; specifically, the ultraviolet radiation (emitted from the black hole) will deposit $\dot{p}_\text{UV} = L/c$ on to the gas, while infrared radiation (re-emitted from dust) will deposit $\dot{p}_\text{IR} = \tau L/c$ on to the gas.  Thus, the total $\dot{p}$ can be approximated as $(1+\tau)L/c \simeq \tau L/c$ for $\tau \gtrsim 1$, which is valid near the peak of AGN activity when the black hole gains most of its mass.  As in DQM11, we set $\tau = 10$.  Lastly, the momentum is returned radially, such that every gas particle within $r_\text{inf}$ receives an equal acceleration.  

In this model, there is no explicit artificial black hole advection algorithm.  Instead, a tracer mass is used to represent the black hole particle; that is, the dynamical mass of the black hole is initialised to $m_\text{BH} = 10^9$ M$_{\astrosun}$ and held constant throughout the simulation.  The black hole particle is now only advected by the gravitational solver; since the tracer mass in the fiducial resolution simulations is $\sim$$10^3m_\text{DM}$ and $\sim$$3\times 10^4m_\text{g}$, it will not undergo artificial dragging by the surrounding particles.  By initialising $m_\text{BH} = 10^9$ M$_{\astrosun}$ rather than $m_\text{BH} = M_\text{BH}$, the mass of each galaxy is increased by 0.01 per cent.  This means that Model DQM has slightly (but unavoidably) different initial conditions than the remainder of the models.  The internal mass of the black hole is still initialised to $10^5$ M$_{\astrosun}$, and this mass allowed to grow as calculated by the accretion rate given in \eqref{Visc}.

Since the dynamical mass of the black hole is fixed, a continual-conditional particle accretion algorithm removes `accreted' gas particles, but does not add their properties to the dynamical mass.  Here, a random particle within $r_\text{inf}$ is removed whenever there is a mismatch between the amount of gas accreted via \eqref{Visc} and the total mass of removed gas particles.

Lastly, black holes merge when they come within $r_\text{inf}$ of each other, regardless of velocity.  In this model, black hole mergers are the only mechanism to increase the dynamical mass of the black hole.

\subsubsection{Model WT: This study}
\label{Mwt}
This model uses the modified Bondi accretion rate given in \eqref{mBondi} for both its simplicity and for its wide use in the literature (e.g. SDH05; \citealp{RHCDHMS06}; \citealp{CDSH09}; \citealp{JNB09}); as in Model SDH, we set $\alpha = 100$.  The energy feedback rate is given by \eqref{Efeed1}, and the feedback energy is equally distributed amongst all the gas particles within $r_\text{inf}$.  This provides an isotropic heating to the core rather than preferentially heating (or super-heating) the particles very near the black hole, which may be transient. 

The black hole advection algorithm is a modified version of that presented in Model ONB.  First, the black hole is displaced towards the centre of mass of the sphere with radius $r_\text{inf}$ centred on the black hole rather than along stellar gradients.  This method still gives preference to the stellar distribution, but also considers the dark matter and gas distributions.  Second, the distance the black hole is displaced has been modified to 
\begin{equation}
\label{deltal5}
\Delta l_\text{WT} = \min(0.10h_\text{BH}, 0.30\left| {\boldsymbol v} \right| \mathrm{d}t, d_\text{CM}),
\end{equation}
where $d_\text{CM}$ is the distance from the black hole to the centre of mass.  We choose to use the black hole smoothing length rather than the softening length since all of the properties near the black hole are calculated using $2h_\text{BH}$ as the characteristic length.  As in \eqref{deltal4}, our coefficients are empirically chosen so that non-negligible displacement would be possible.  Specifically, for the first option, $h_\text{BH} < \epsilon_\text{S2}$, so its coefficient needs to be increased; for the second option, our resolution is higher than in ONB08, thus we will have a smaller $\mathrm{d}t$, thus we again need a larger coefficient.  We choose this method rather than potential-well method of Models SDH and BS for the following reasons: If the AGN feedback creates a gas void around the black hole, then either there will be no artificial displacement (allowing for the possibility of unnatural motion) or the black hole will be coupled to a gas particle on the edge of the void.  In this method, as in Model ONB, the black hole only moves towards a particle rather than being coupled to it. 

As in Models ONB and DQM, we include a continual-conditional particle accretion algorithm:  When $M_\text{BH} > m_\text{BH} + m_\text{g}/2$, we accrete the gas particle that is nearest to the black hole.  The term $m_\text{g}/2$ forces the dynamical and internal mass to oscillate around one another, and choosing the nearest particle is to reduce random effects in the simulation.

The black hole merger algorithm is the same as given in Model SDH.  We choose this algorithm since it had a distance and a velocity criteria, and its velocity requirement is less stringent than in Models BS and ONB.  The merger algorithms in Models BS and ONB were created for cosmological models, whose resolution is lower than our fiducial resolution runs, thus we argue that our implementation of their merger prescriptions are more stringent than they intended.  

\subsubsection{Additional models}
\label{Madd}

We have tested four additional models, each of which is a slight variation of one the above models.  These models are identical to their parent models described above, except for the variation listed below.  
\begin{itemize}
\item \emph{Model BSw}:  This uses the black hole advection algorithm of Model WT; early tests of Model BS showed very erratic black hole motion that could possibly compromise the results.
\item \emph{Model ONBc}: This uses a very conservative search algorithm to calculate $\dot{M}_\text{drag}$; this yields a very small $R_\text{RSF}$ and a very low accretion rate.
\item \emph{Model WTh}: For a resolution test, this model uses $h_\text{min} \rightarrow h_\text{min}/2$; this should only impact calculations performed in very dense regions.  
\item \emph{Model DQMe}: This model uses $r_\text{inf} \equiv 4 \epsilon_\text{S2}  = 1.17$ kpc (compared to $r_\text{inf} \sim 73$ pc of Model DQM); in DQM11, they fix $r_\text{inf} \equiv 4 \epsilon_\text{S2}$, although this value is $\sim$188 pc in their models.
\end{itemize}

\section{Results}
\label{results}
Each of our models was evolved through a merger event, similar to that of SDH05.  Model ONB was evolved for 1.1 Gyr and the rest were evolved for 1.5 Gyr (including Model ONB$_\text{l}$).  By returning the feedback energy to the halo gas in Model ONB, a dense galactic core formed near the core merger epoch, which resulted in an extremely large wall-clock time per step for the fiducial resolution version.  

\begin{figure*}
\begin{center}
\includegraphics[width=1.0\textwidth]{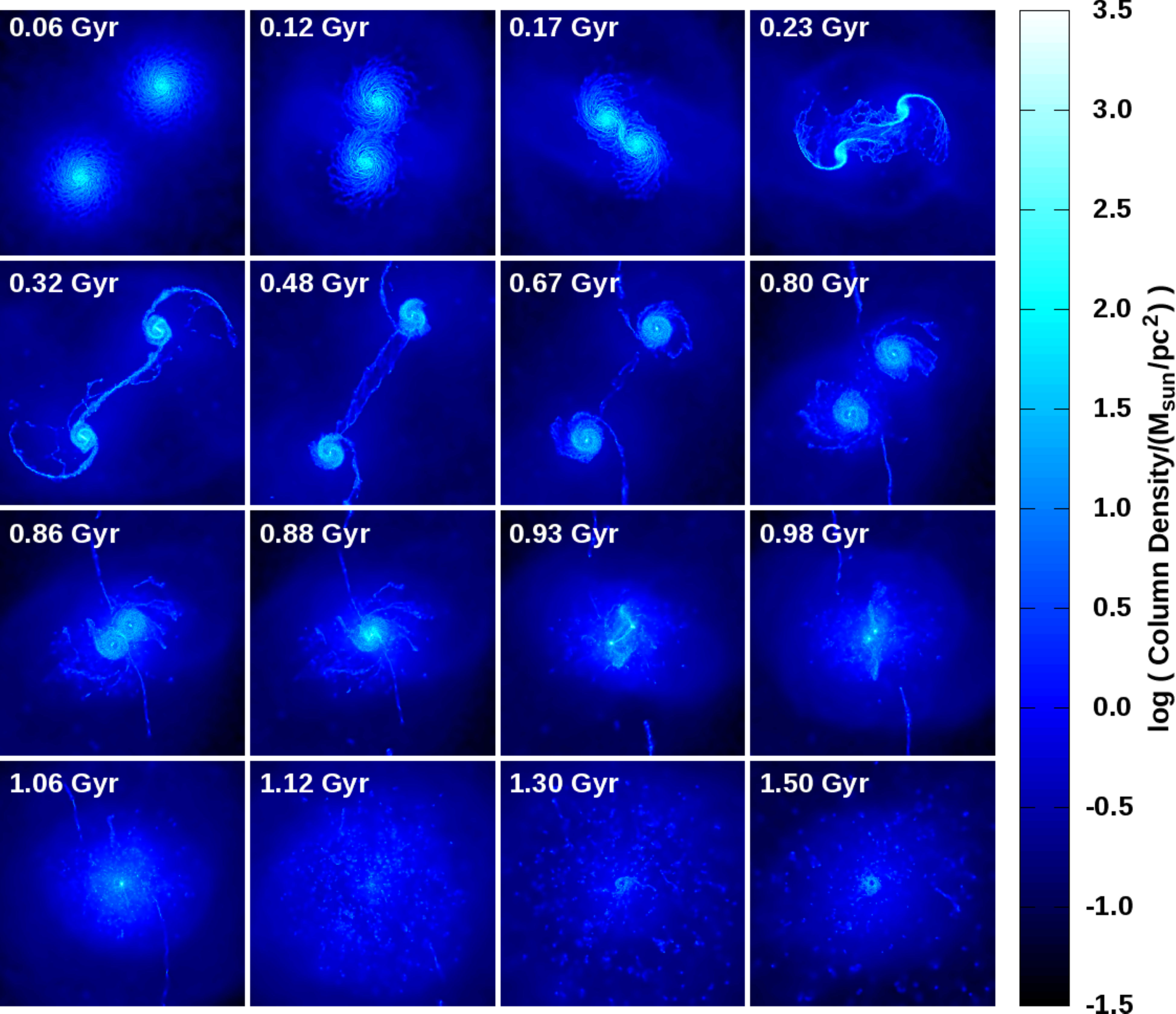}
\end{center}
\caption{Evolution of gas column density for Model WT.  Times from the onset of the simulation are listed in each frame; each frame measures 100 kpc per side, with an image resolution of 98 pc pixel$^{-1}$.}\label{r_m1snap}  
\end{figure*}

\begin{figure*}
\begin{center}
\includegraphics[width=1.0\textwidth]{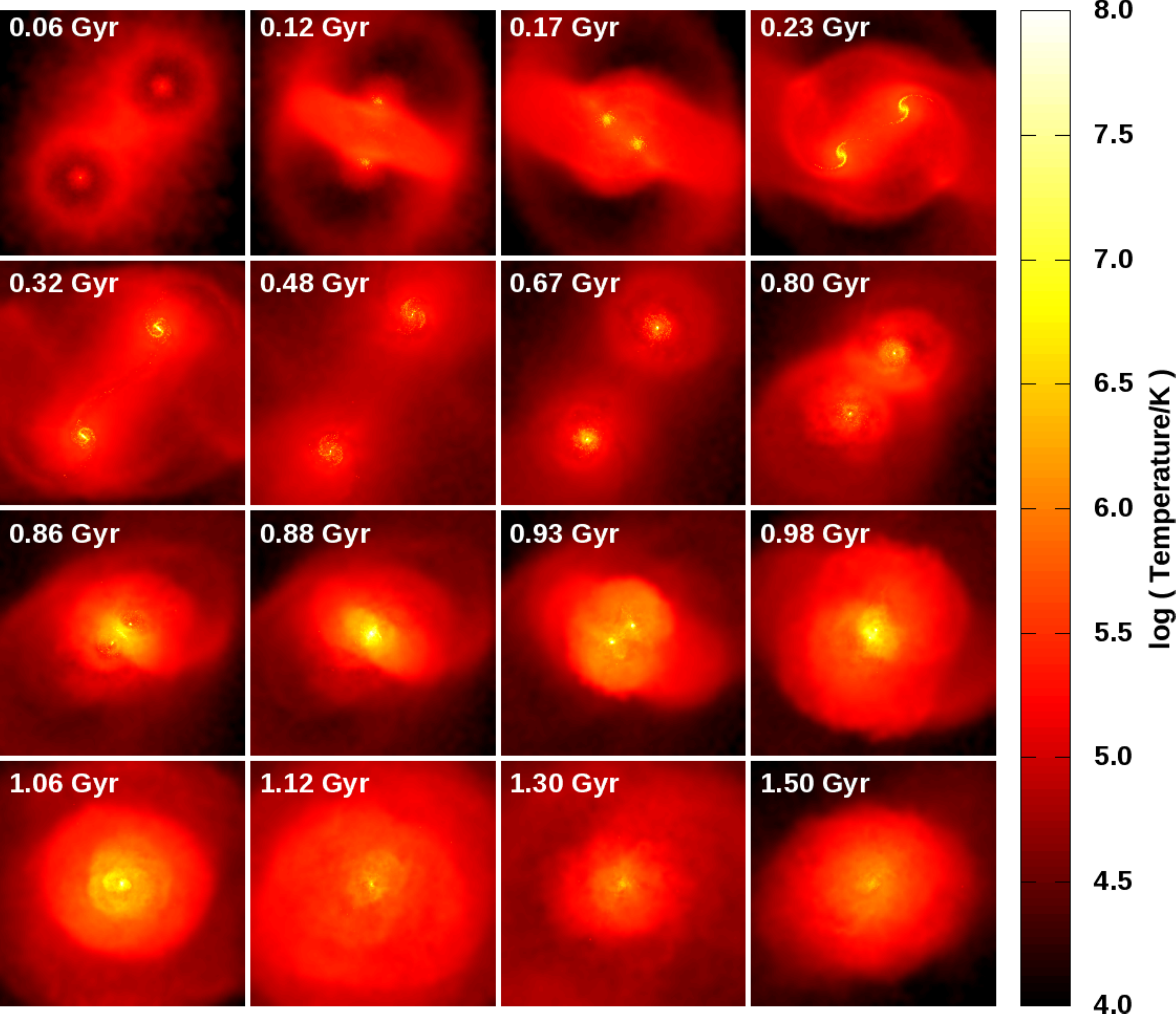}
\end{center}
\caption{Evolution of gas temperature for Model WT.  Times from the onset of the simulation are listed in each frame; each frame measures 100 kpc per side, with an image resolution of 98 pc pixel$^{-1}$.  The top right panel at 0.23 Gyr has a large filling factor of supernova events, caused by tidal interactions at first periapsis.  Although ubiquitously occurring through the galaxies, the global star formation rate is at a local minima (see section \ref{r_gsfr}).} \label{r_m1snapT}  
\end{figure*}

Each model followed a similar qualitative history, which is shown for Model WT in Figs. \ref{r_m1snap} and \ref{r_m1snapT} for the gas column density and gas temperature, respectively.
From this evolution, we see four significant epochs: first periapsis, apoapsis, second periapsis, and core merger.  The times of these epochs are calculated using the black holes as proxies for the centres of the galaxies, thus the epochs essentially represent the local minimum and maximum separation of the two black holes.  In all models, first periapsis occurs at 166 Myr and apoapsis occurs at 480 Myr; the latter is sustained for approximately 100 Myr.  Second periapsis occurs approximately 11 Myr earlier for Model DQM than for the rest of the models.  We have verified that this more rapid evolution is due to the inclusion of a more massive sink particle representing the black hole.  In two-body simulations, where each particle represents one galaxy and dynamical friction is avoided, the period of the particles representing galaxies in Model DQM is 3.4 Myr (0.23 per cent) shorter than the particles representing the galaxies without tracer masses. For the full system, we find that the maximum separation at apoapsis is very slightly smaller (1.1 per cent), likely due to slightly higher dynamical friction, leading to an earlier second periapsis.  Lastly, the core merger takes approximately 10 Myr to complete, and the onset of this process spans a range of 15 Myr amongst the different models.  See Table \ref{r_imptime} for a list of when these epochs occur.    

\begin{center}
\begin{table}
{\small
\hfill{}
\begin{tabular}{l|r r}
\hline
                        & Dynamic Mass   	& Tracer Mass   \\
\hline
\hline
First Periapsis/Gyr	& 0.166 (0.166)		& 0.166 (0.166) \\
Apoapsis/Gyr		& 0.480	(0.480)		& 0.480 (0.480) \\
Second Periapsis/Gyr	& 0.884	(0.870)		& 0.872 (0.861) \\
Core Merger/Gyr	    	& 0.987	(0.973)		& 0.970 (0.962) \\
\hline
\end{tabular}}
\hfill{}
\caption{Important epochs from the onset of the simulation for the fiducial (low) resolution models.  Average times are listed when a range is available.  The Dynamic Mass models are Models SDH, BS, ONB and WT; the Tracer Mass model is Model DQM.}
\label{r_imptime} 
\end{table}
\end{center}

\begin{figure*}
\begin{center}
\includegraphics[width=0.24\textwidth]{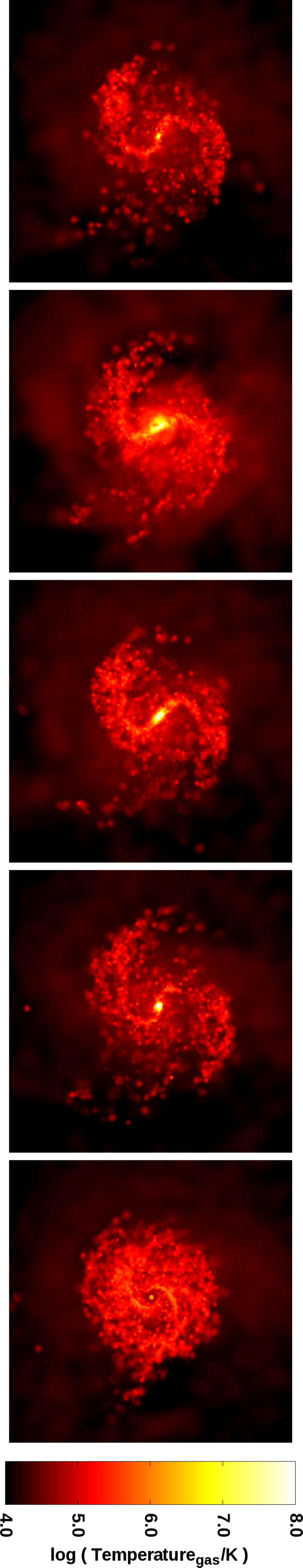}
\includegraphics[width=0.24\textwidth]{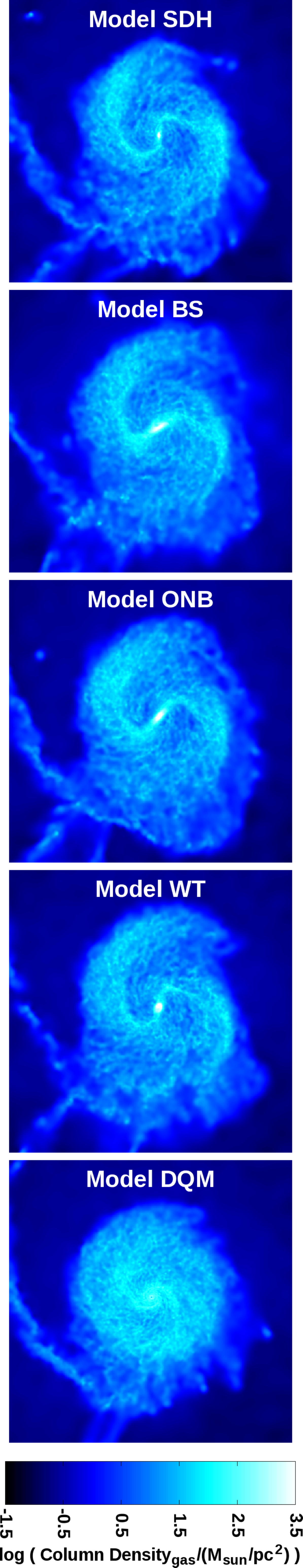}
\includegraphics[width=0.24\textwidth]{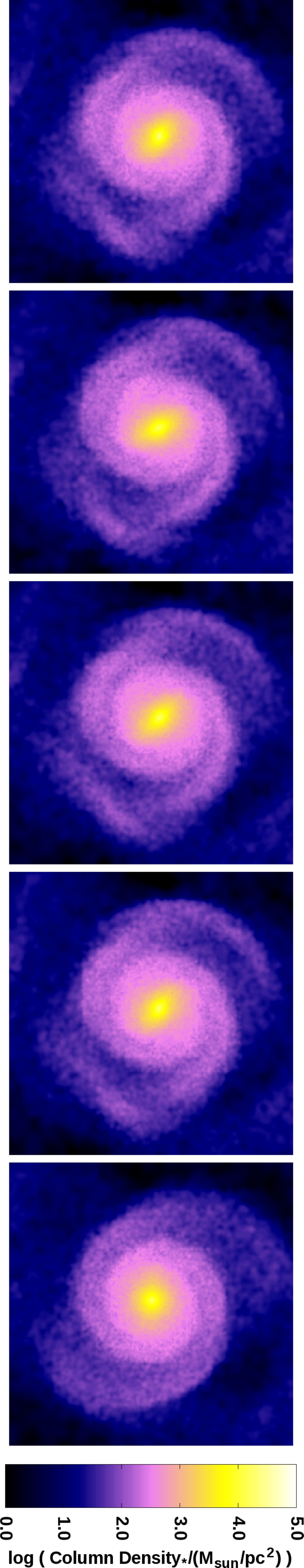}
\end{center}
\caption{Top left galaxy of each model at apoapsis.  Varying morphologies for the different models emerge as the galactic discs reform after first periapsis.  The lack of bar in Model DQM is a result of the more massive black hole particle (i.e. the tracer mass).  Each frame measures 20 kpc per side, with an image resolution of 39 pc pixel$^{-1}$.  \emph{From left to right}: Gas temperature, gas column density and stellar column density.}\label{r_all}  
\end{figure*}

\begin{figure}
\begin{center}
\includegraphics[width=1.0\columnwidth]{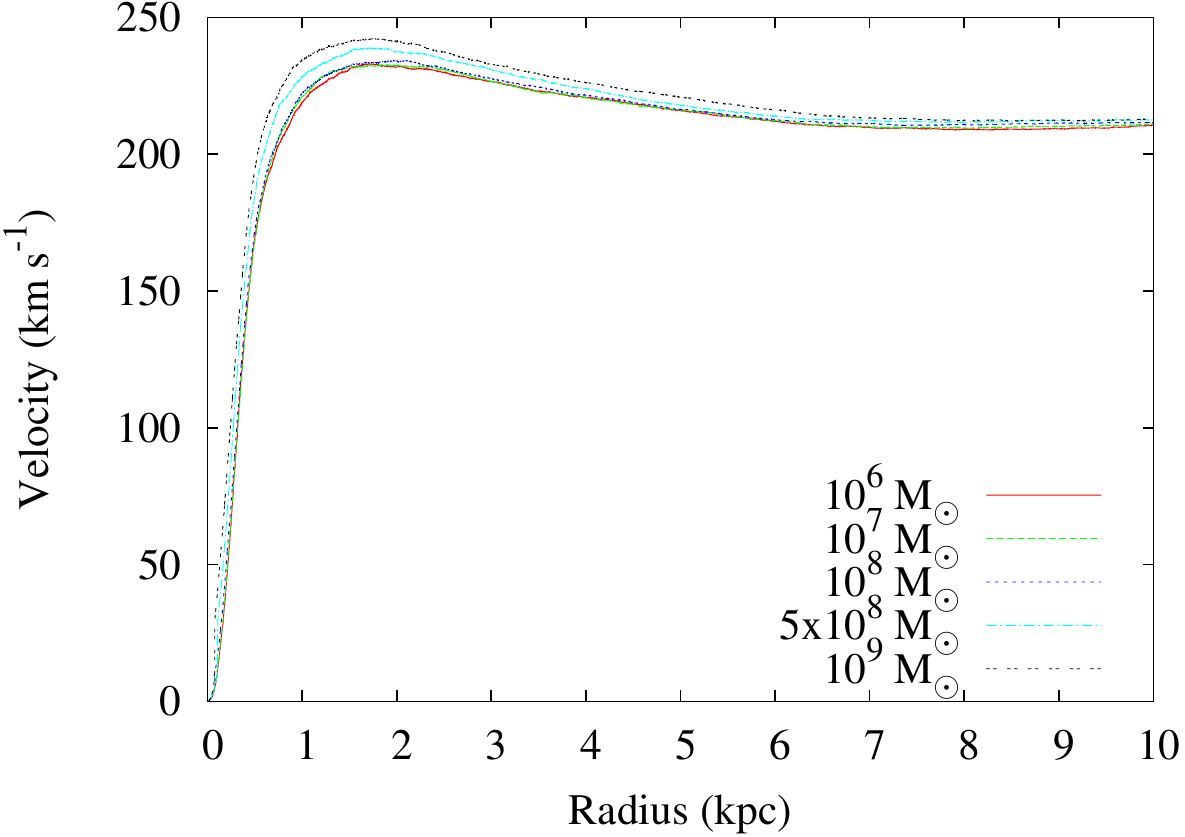}
\end{center}
\caption{The rotation curves for five low resolution test models, where the mass of the tracer mass has been varied.  For each model, the rotation curve is for the top left galaxy, and is calculated at 0.25 Gyr just prior to the onset of bar formation.  The parent model is Model WT$_\text{l}$.}\label{r_rotvel}  
\end{figure}

As might be expected, there are notable morphological differences between the models; this is readily apparent in Fig. \ref{r_all}, which displays the gas temperature, gas column density and stellar column density of the top left galaxy in each model at apoapsis.  By apoapsis, a bar has developed in our non-tracer mass models (Models SDH, BS, ONB and WT), with bar strength depending on model.  We have verified that the lack of bar formation in Model DQM is a result of the additional mass from the tracer mass and not from the kinetic feedback.  This extra mass causes an increase in the rotational velocity curve of the galaxy.  See Figure \ref{r_rotvel} where we have plotted the (S2 softened) rotation curve for five variations on Model WT$_\text{l}$ just prior to the onset of bar formation; in this test suite, a tracer mass represents the black hole and has been varied between $10^6$ and $10^9$ M$_{\astrosun}$.  The rotation curves for $M_\text{tracer} = 10^{6-8}$ M$_{\astrosun}$ are similar, and the peak velocity is 2.5 per cent higher for $M_\text{tracer} = 5\times10^8$ M$_{\astrosun}$ and 4.0 per cent higher for $M_\text{tracer} = 10^9$ M$_{\astrosun}$.  The higher rotational velocities for $M_\text{tracer} = 10^9$ M$_{\astrosun}$ stabalises the galaxy against bar formation.  


To ensure that our initial black hole mass is reasonable, we ran Model WT$_\text{l}$ with seed black hole masses of $10^5$, $10^6$ and $10^7$ M$_{\astrosun}$.  All three models had final black hole masses within 1.5 per cent of one another, and all three remnants had $M_\text{BH}$--$\sigma$ relations that agreed with the observed relation within the one-sigma standard deviation.  

\subsection{Black hole advection}
\label{R_bha}

\begin{figure}
\begin{center}
\includegraphics[width=1.0\columnwidth]{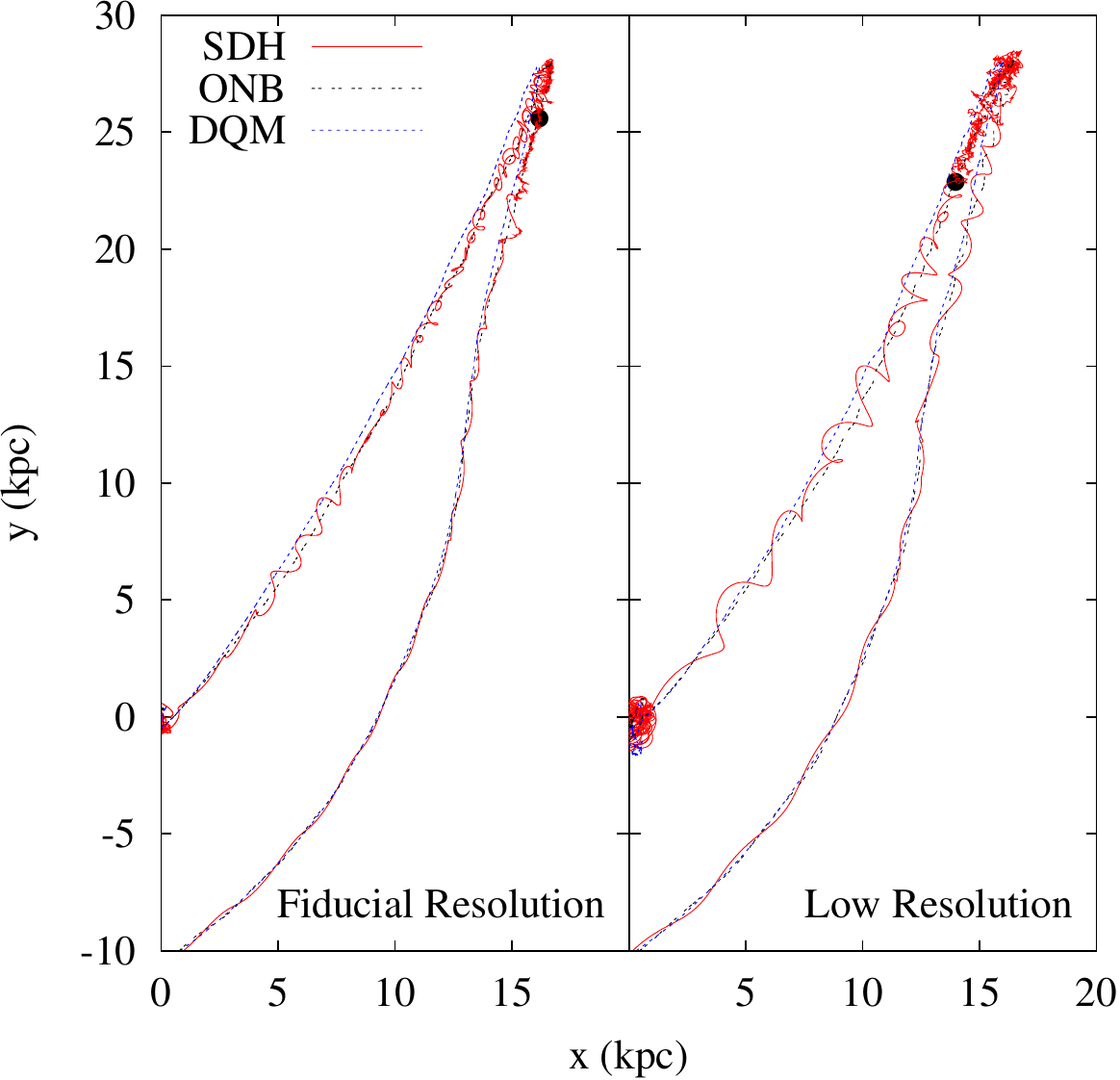}
\end{center}
\caption{A portion of the unaveraged black hole path (including the turnaround point at apoapsis) in Models SDH, ONB and DQM for the fiducial (left) and low resolution (right) simulations.  The dot indicates where the black hole advection algorithm turns off for Model SDH.  The origin of the system is $(x,y,z) = (0,0,0)$ and the black hole originates at $(-27.6, -21.9, 0.0)$.}\label{r_bhloc} 
\end{figure}

The location of the black hole with respect to the galactic core plays a fundamental roll in determining the gas properties used to calculate the accretion and energy feedback rates.  Even small displacements with respect to the centre of the potential well can have notable effects.  We have studied four artificial advection algorithms, each yielding different results.  Since the results presented in the rest of this paper are implicitly coupled to the behaviour of the advection algorithm, we discuss this issue first.  See Fig. \ref{r_bhloc}, where we have isolated the path of one black hole in Models SDH, ONB and DQM.  

Without any artificial advection, low mass black hole particles are easily dragged by the local particles (whose mass may be comparable or greater than the mass of the black hole); even with artificial advection, some black hole `chaotic' motions (or oscillations) persist, depending on the algorithm.  The smoothest path is from the tracer mass (Model DQM); however, the increased mass of the galaxy slightly adjusts its trajectory, decreases its evolution time compared to the rest of the models, and impacts the disc morphology (i.e. prevents bar formation).  The least smooth path is from coupling the black hole to the gas particle with the minimum potential energy, provided that $v_\text{rel} < 0.25c_\text{s}$ (Models SDH and BS).  Frequently, there is no gas particle satisfying the velocity criterion, thus the black hole oscillates about the galactic core.  When there is a gas particle satisfying the criteria, it is likely near the periphery of $r_\text{inf}$ since the inner particles are typically too hot.  Thus, the stringent velocity criterion permits both dragging by two-body forces and large artificial jumps in position.  In Models SDH and BS, the artificial advection is turned off when $m_\text{BH} > 10m_\text{g}$; in the fiducial (low) resolution, the motions become smaller (larger) and more chaotic.  We note that SDH05 and BS09 both use a multiphase star formation model, which produces a smoother temperature profile around the black hole than our star formation model.  This difference may explain the poor advection in Models SDH and BS.

From the simulations, the advection methods in Models ONB and WT appear to have some advantages; they do not add mass to the galaxy as in the tracer mass method, nor do they couple to gas particles.  By associating the direction of motion to stellar densities (Model ONB) or the local centre of mass within $r_\text{inf}$ (Model WT), undo influence of gas particles, which are dynamically and continuously affected by the AGN feedback, is avoided.  Lastly, by limiting the distance a black hole can be artificially advected per iteration, it is possible for it to remain in the centre of a void.

The advection algorithms in Models DQM, ONB and WT are comparatively unaffected by resolution; this is expected since the tracer mass is independent of the local gas particles, and local averages vary little with resolution.  The advection algorithm in Models SDH and BS are more resolution dependent; with lower resolution comes a larger $r_\text{inf}$, thus the black hole has a larger distance it can be artificially displaced.  It is worth noting that Fig. \ref{r_bhloc} clearly shows that the oscillations of Model SDH is damped with higher resolution, but is not completely removed.  

Due to the notable black hole motion in Model BS, we ran a variation, Model BSw, which uses the black hole advection algorithm of Model WT.  In this model, the black hole had smoother accretion rates and more continuous gas densities and temperatures near the black hole.  The outbursts near apoapsis were stronger and more symmetric between the two galaxies than in Model BS.  Ironically, although we might argue that Model BSw is inherently better than Model BS because of the better black hole trajectories, both models yield remnants with similar total black hole masses and stellar velocity dispersions, even though the black holes in Model BS never merge and the feedback history of the two models is different.  

\subsection{Accretion rate}
\label{R_ar}

\begin{figure}
\begin{center}
\includegraphics[width=1.0\columnwidth]{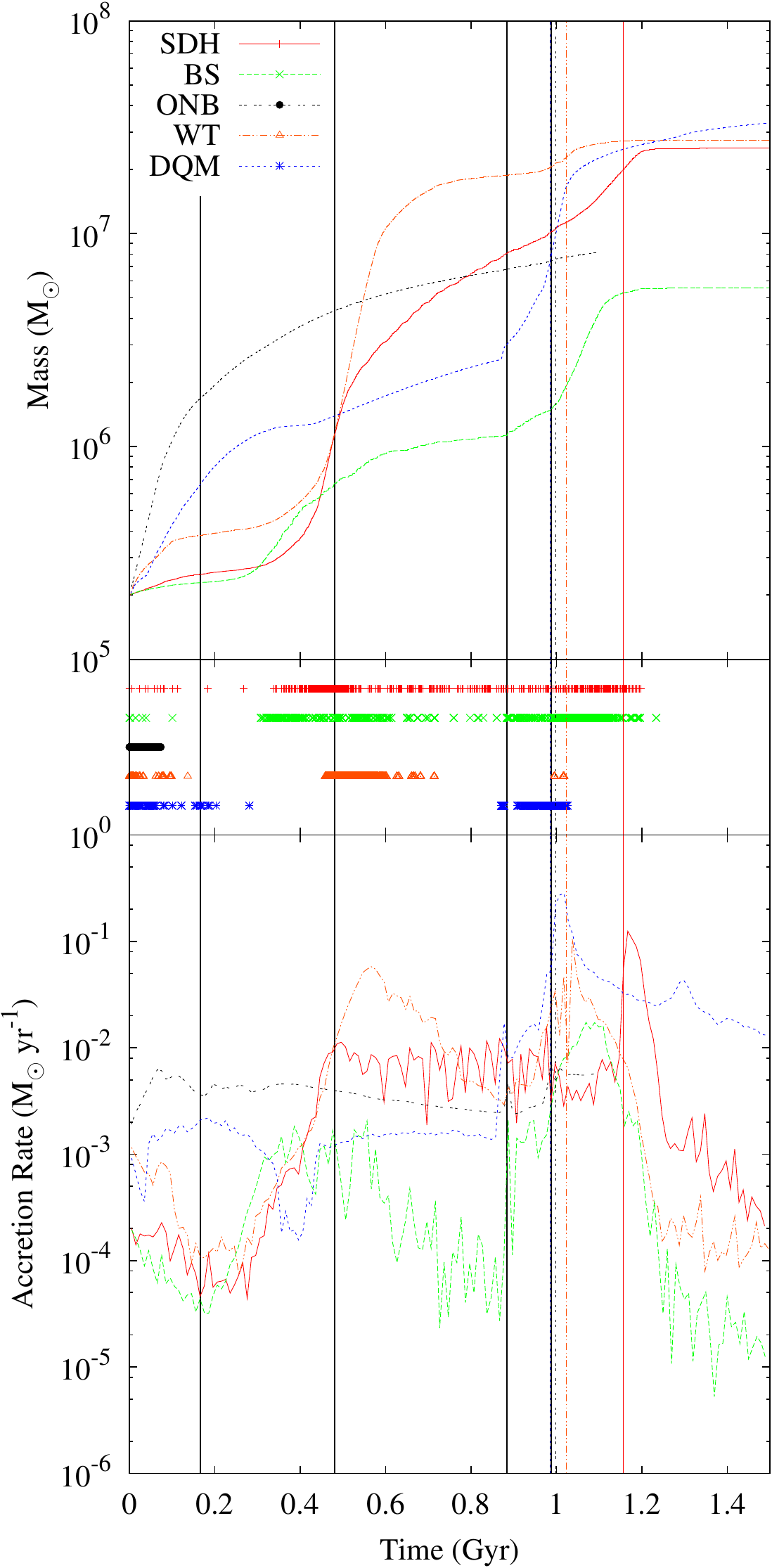}
\end{center}
\caption{\emph{Bottom}: The accretion rates for each of the fiducial resolution models geometrically averaged over both black holes and plotted in bins of 10 Myr.  \emph{Top}: The total black hole mass in each fiducial resolution simulation.  \emph{Middle}: Points represent when a black hole is accreting at $\dot{M}_\text{Edd}$.  The vertical line of the same linestyle indicates when the black holes merge.  The four black vertical lines indicate the time of first periapsis, apoapsis, second periapsis and core merger.}\label{r_bhaccretionmass}  
\end{figure}

The mass evolution of the black hole is a key aspect of the AGN feedback models and is an important diagnostic since it sets the $M_\text{BH}$--$\sigma$ results.  In the top panel of Fig. \ref{r_bhaccretionmass}, we show the total black hole mass over time; in the bottom panel, we show the accretion rates (geometrically averaged over both black holes and plotted in bins of 10 Myr); in the middle panel, we show when a black hole is accreting at its Eddington limit.  
There are three major epochs of black hole growth: at the beginning (Models ONB, WT and DQM), at apoapsis (Models SDH, BS and WT), and at core merger (Models SDH, BS and DQM).  It is typically at these times when a black hole is undergoing Eddington accretion; see Table \ref{r_accretetion_timeEL} for the percent of time that a black hole is accreting at its Eddington limit.
\begin{center}
\begin{table}
{\small
\hfill{}
\begin{tabular}{l r r r r}
\hline
           & Pre-BH Merger & Merged BH \\
\hline
\hline
Model SDH  &  8.32 ( 6.82) &  1.95 (0.04)\\  
Model BS   &  5.87 ( 2.26) &  ---  ( ---)\\
Model ONB  &  5.78 ( 3.17) &  0.00 (0.00)\\ 
Model WT   &  9.99 ( 5.75) &  0.00 (0.00)\\
Model DQM  &  2.35 ( 8.22) &  6.86 (1.42)\\
\hline
Model BSw  &  2.11 ( 1.18) &  0.00 (0.00)\\
Model ONBc &  0.00 ( 0.37) &  0.00 (0.00)\\  
Model WTh  &  8.78 ( 6.94) &  2.10 (0.00)\\
Model DQMe & 17.72 (15.74) &  1.20 (0.00)\\
\hline
\end{tabular}}
\hfill{}
\caption{Percent of time black holes (BHs) are accreting at their Eddington limit; the pre-BH merger times are averaged over both black holes.  Dashed lines indicate that the black holes did not merge prior to 1.5 Gyr.  The percentages for fiducial resolution Models ONB and ONBc are taken at 1.1 Gyr and 1.25 Gyr, respectively.}
\label{r_accretetion_timeEL} 
\end{table}
\end{center}

Three models use the Eddington-limited Bondi accretion rate: Models SDH, BS and WT.  In SDH05 (see bottom panel of their fig. 14), the accretion rate increases from the onset of the simulation until prior to second periapsis, at which point there is only a slight decrease.  At core merger, there is an increase of approximately 1.8 dex, followed by a sudden drop as the system is totally disrupted.  Since our initial conditions and our Model SDH was constructed to mimic SDH05, we expect similar results, which, within reason, are obtained.  In both Models SDH and WT, there is an initial decline in accretion rate as the galaxies are relaxing; this initial relaxation does not occur in SDH05 since the vertical structure of the gas is set in hydrostatic equilibrium, thus this initial difference is expected.  After first periapsis, gas is funnelled into the core, thus, in agreement with SDH05, there is an increasing accretion rate.  At apoapsis, there is a plateau in Model SDH, but a short-lived spike in Model WT.  At core merger, there is an increase followed by a rapid decrease as the systems are totally disrupted.

\begin{figure}
\begin{center}
\includegraphics[width=1.0\columnwidth]{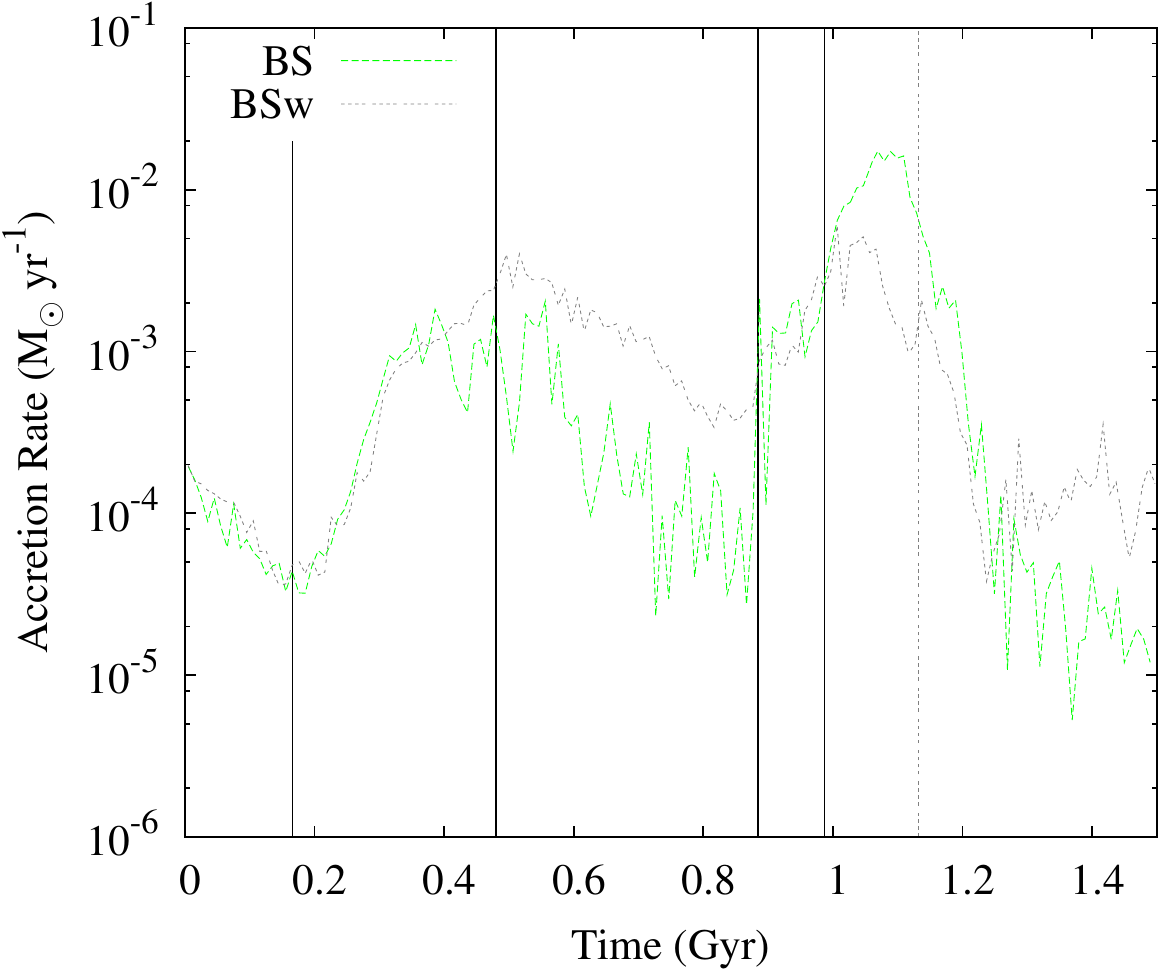}
\end{center}
\caption{The accretion rate for Models BS and BSw, geometrically averaged over both black hols and plotted in bins of 10 Myr.  The vertical line of the same linestyle indicates when the black holes merge.  The four black vertical lines indicate the time of first periapsis, apoapsis, second periapsis and core merger.}\label{r_accretionBS}  
\end{figure}

For the first 400 Myr, the accretion rate of Model BS follows the same trends as Models SDH and WT.  During this time, $\alpha(n_\text{H}) > 100$, which explains why $\dot{M}_\text{Model BS} > \dot{M}_\text{Models SDH \& WT}$ for $230 \lesssim t/\text{Myr} \lesssim 400$.  After this point, $\dot{M}_\text{Model BS}$ decreases, which is a result of the strong outflows from feedback and the motion of the black holes; the latter prevents the black holes from remaining in a steady environment, resulting in rapidly varying gas characteristics around the black hole.  The accretion rates in Model BSw continue to increase until apoapsis, after which they decrease due to strong outflows, but to a smoother accretion profile that is slightly higher than in Model BS; see Fig. \ref{r_accretionBS}.

In DQM11, there is a large jump in $\dot{M}_\text{visc}$ followed by a slow decline shortly after first periapsis and again starting at second periapsis; see the bottom panel of their fig. 1.  Their initial galactic separation is approximately twice of ours, thus their evolution times are longer, which may account for differences between our models and theirs.  Also, their mass and spatial resolution are slightly better than in the models presented here, thus their feedback is likely distributed amongst fewer particles, which keeps their void relatively smaller (see discussion in \S \ref{R_efa}).  

In Model DQM, there is not the accretion epoch between first periapsis and apoapsis that occurs in DQM11; this is a result of both a shorter time during which gas can fall into the galactic cores, and the slightly larger $r_\text{inf}$ which allows for efficient feedback to maintain the void during non-cataclysmic events.  At second periapsis, there is an increase of approximately 2.2 dex in Model DQM compared to DQM11's increase of approximately 3 dex.  After the merger, both accretion rates drop off, but DQM11 falls off faster.  

ONB08 is a cosmological simulation run from $z = 49$ to $z = 0$.  Over the final 7 Gyr of their simulation, there are no significant galaxy mergers and there is negligible black hole growth (see their fig. 8).  During this time, their accretion rate lies comfortable within the RAIF regime by more than an order of magnitude.  In Model ONBc, there is a low accretion rate that is continually decreasing and comfortably within the RAIF regime; however, this model uses a very conservative search algorithm to calculate $\dot{M}_\text{drag}$, thus the low accretion rate is likely artificially low because of the numerical scheme.  When we use a less stringent search algorithm that is better suited for use in {\sc Hydra} at the resolutions we present, the accretion rate quickly rises to $\dot{M}_\text{drag} \sim 0.005$ M$_{\astrosun}$yr$^{-1}$ and remains approximately constant for the remainder of the simulation.  This relatively constant rate is a result of the feedback energy being deposited into the halo gas rather than modifying the galactic core.  Since the feedback is being returned to the halo gas, there are negligible core differences between Models ONB and ONBc, except for the accretion rate.

\subsection{Energy feedback algorithm}
\label{R_efa}
\begin{figure}
\begin{center}
\includegraphics[width=1.0\columnwidth]{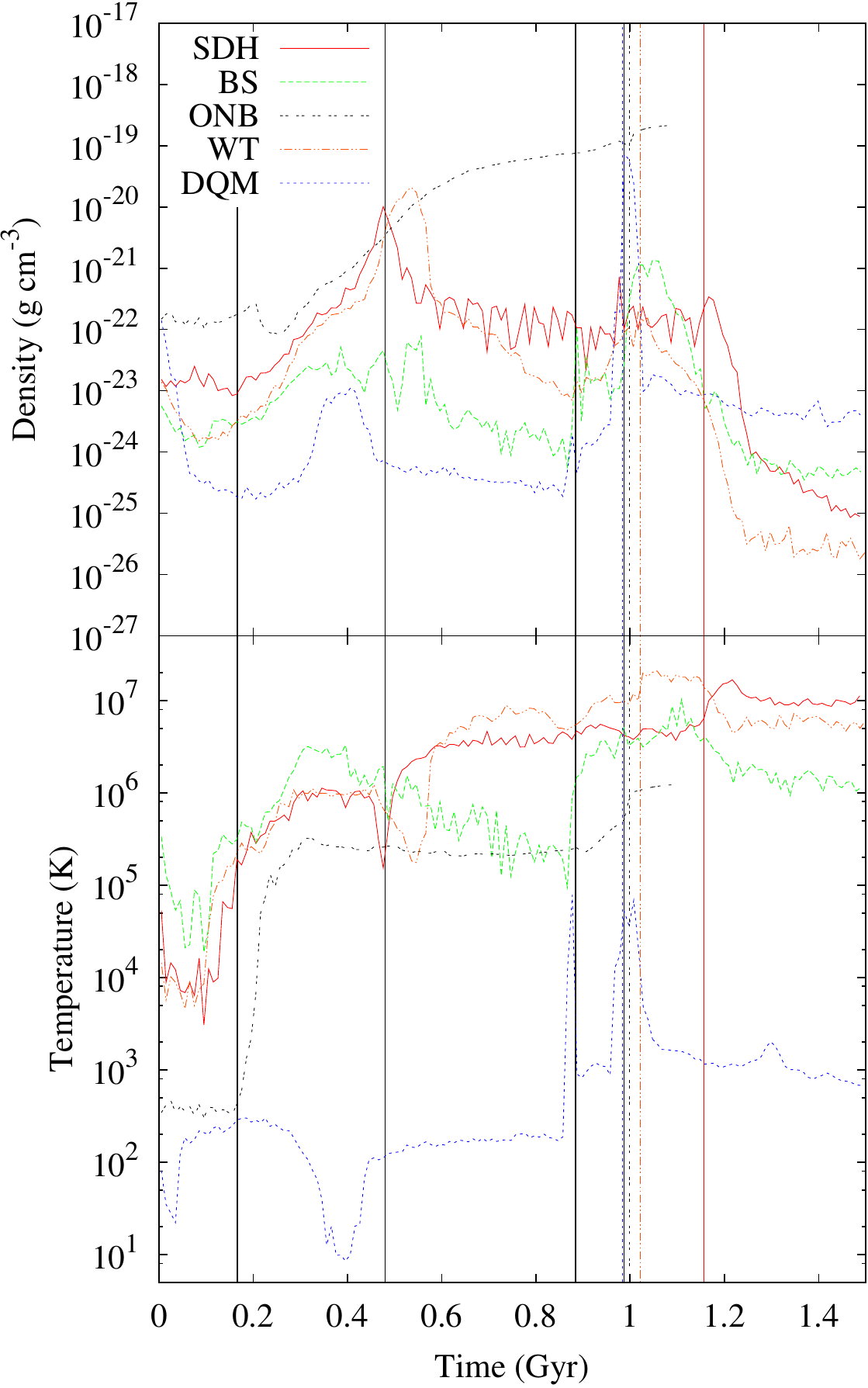}
\end{center}
\caption{Gas density (top) and temperature (bottom) within $r_\text{inf}$ the black holes, geometrically averaged over both black holes and plotted in bins of 10 Myr.  The vertical line of the same linestyle indicates when the black holes merge.}\label{r_rhotemp} 
\end{figure}

The two broad categories of returning the feedback energy are thermally, $\dot{E}$, and kinetically, $\dot{p}$.  Thermal energy is returned to gas particles in Models SDH, BS, ONB and WT by increasing the internal energy of the particles; kinetic energy is returned to gas particles in Model DQM by increasing the momentum of the particles.  Although these two approaches are related by $\dot{E} = \dot{p}c$, both forms of feedback yield drastically different results, which are readily apparent on both large scales and small scales (e.g. see Fig. \ref{r_rhotemp} for gas density and temperatures around the black holes).  

The feedback in Model ONB is returned to the gas particles with $\rho < 5\times 10^{-26}$ g cm$^{-3}$; thus, this energy is distributed to halo gas particles with distances 0.5 $\lesssim r/\text{kpc} \lesssim$ 3.5 ($7 \lesssim r/r_\text{inf} \lesssim 48$) from the black holes, which leaves the galactic cores relatively unmodified by AGN feedback.  As gas falls towards the galactic cores (due to tidal interactions at first periapsis), it remains there since the AGN feedback is not local enough to remove it, hence these models have the highest core density.  The major core heating events in this case are from shock heating of the infalling gas after first periapsis and from the final core collisions.

The AGN feedback in the remaining models is delivered directly to the core region, leading to a more distinct modification of the entire system.  From the onset of the remaining thermal feedback simulations, there is a higher core temperature and a lower core density than in Model ONB.  Tidal interactions at first periapsis cause gas to be funnelled towards the cores, but unlike Model ONB, the temperature continues to rise from AGN heating, and the density remains moderated.  Shortly after first periapsis, a galactic bar starts to form and persists strongly for approximately 200 Myr; as it dissipates there is a nuclear inflow of gas, accounting for the local density maxima near apoapsis in Models SDH and WT.  After the dissipation of the bar, the galaxies begin to eject gas, preferentially along the polar axis.  As the cores merge, there is a final set of feedback events as the gas is heated and blown away; after this, accretion rates and core densities drop while core temperature remains approximately constant.  

Kinetic feedback affects the galactic core in different quantitative and qualitative ways than thermal feedback.  In this model, the energy is returned radially to gas particles as a momentum boost, which quickly and efficiently forms a void around the black holes; in Model DQM, a void of $r \sim 0.55r_\text{inf}$ is formed within the first 60 Myr.  Once the void is formed, it persists for the remainder of the simulation.  Strong events, such as the infalling gas after first periapsis or the collision at core merger, can briefly decrease the void radius, increasing the core density.   Otherwise, the void and core characteristics are held constant, as we can see by an approximately constant temperature and density profile between first periapsis and core merger.  

Model DQMe uses the same kinetic feedback algorithm as in Model DQM, but instead $r_\text{inf} \equiv 4\epsilon_\text{S2}$, which leads to an initially higher accretion rate.  The increased amount of feedback energy is returned to more particles, delaying the formation of the void.  When the void forms after 140 Myr, it has a radius of $r \sim 0.85r_\text{inf} \sim 1$ kpc, which is highly unrealistic.  This void is more rigid than the void in Model DQM, only being affected at second periapsis and core merger.  The unrealistically large $r_\text{inf}$ in Model DQMe also leads to a premature merger of the black holes, which is poorly handled by the merger subroutine.  Thus, the importance of choosing a physical $r_\text{inf}$ is clear: catastrophic and unphysical results will ensue if it is not chosen appropriately.

\subsection{Particle accretion algorithm}
\label{R_app}

\begin{figure}
\begin{center}
\includegraphics[width=1.0\columnwidth]{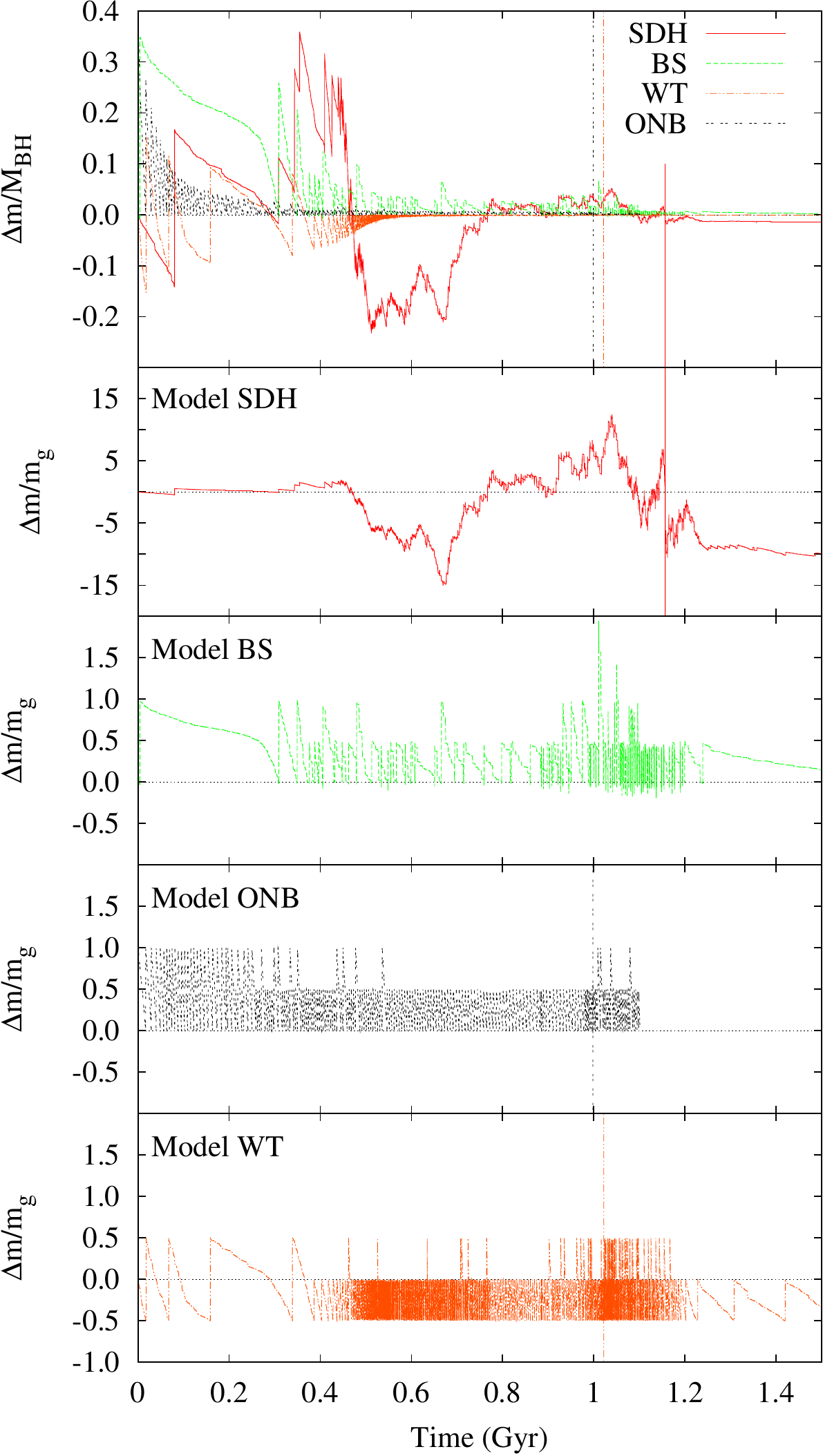}
\end{center}
\caption{A comparison of internal mass, $M_\text{BH}$, and dynamical mass, $m_\text{BH}$, for one black hole in selected fiducial runs.  \emph{Top}: The ratio $\Delta m / M_\text{BH}$, where $\Delta m = m_\text{BH} - M_\text{BH}$.  \emph{Bottom Four Panels (top to bottom)}:  The ratio $\Delta m/m_\text{g}$ for Model SDH (stochastic-unconditional), Model BS (stochastic-conditional), Model ONB (continual-conditional) and Model WT (continual-conditional particle accretion).  The vertical line of the same linestyle in the top panel indicates when the black holes merge.}\label{r_bhdmass} 
\end{figure}

A particle accretion algorithm should accrete particles such that the internal mass, $M_\text{BH}$, and dynamical mass, $m_\text{BH}$, of any given black hole remain similar for all time.  In Model DQM, a tracer mass is used to represent the black hole particle, rendering the comparison between internal and dynamic masses moot; these models use a continual-conditional algorithm to remove `accreted' gas, thus any conclusions we make regarding this category can be applied.  In the remainder of our fiducial resolution models, the first accretion event will increase the dynamical mass by 36 per cent since $m_\text{BH} = 0.36m_\text{g}$.  All subsequent accretion events will add relatively less mass to the black hole particle.  Thus we expect the ratio $\Delta m / M_\text{BH}$, where $\Delta m = m_\text{BH} - M_\text{BH}$, to start at $|\Delta m / M_\text{BH}| \lesssim 0.36$ and decay to zero with time.  In the top panel of Fig. \ref{r_bhdmass}, we plot $\Delta m / M_\text{BH}$ for one black hole in Models SDH, BS, ONB and WT.
The expected behaviour is obtained in Models BS, ONB and WT; the decline in Models ONB and WT is smoother than in Model BS since there is no stochastic component to the accretion.  The expected decline is not observed in Model SDH, where the ratio has no apparent trend, and at late times $\Delta m / M_\text{BH} \rightarrow -0.01$ rather than zero.  

The black holes do not all grow at the same rate.  Thus, for a normalised comparison, we plot $\Delta m / m_\text{g}$ in the bottom four panels of Fig. \ref{r_bhdmass} for Models SDH, BS, ONB and WT, respectively.  Since every accretion event adds mass $m_\text{g}$ or $m_\text{g}/2$ to the dynamical mass, we expect $|\Delta m/m_\text{g}| \le 1$ for all time, unless multiple particles are accreted in one step, which can occur in the stochastic models.  

In Model SDH, particle accretion events are only dependent on the environment.  This lack of $M_\text{BH}$--$m_\text{BH}$ coupling permits the internal and dynamic masses to follow different histories, as seen by the lack of trend in the second panel of Fig. \ref{r_bhdmass}.  The discontinuity in $\Delta m / m_\text{g}$ at core merger is a result of both black holes following different mass growth histories, both internally and dynamically.  

The stochastic-conditional accretion algorithm in Model BS allows for fluctuations in an otherwise smooth process, and the probability's dependence on $\Delta m$ forces good agreement between the internal and dynamic masses.  Finally, by design, Model ONB (WT) always maintains $0.0 \le \Delta m/m_\text{g} \le 1.0$ ($-0.5 \le \Delta m/m_\text{g} \le 0.5$), displaying the effectiveness of the continual-conditional particle accretion methods.  

\subsection{Black hole merger algorithm}
\label{R_bhma}

All of the models include a black hole merger prescription, which conserves internal, dynamical and tracer mass; see Table \ref{r_mergertime} for precise merger times.  Thus, the merger should have limited gravitational impact on the surrounding environment.  However, the amount of feedback energy returned pre- and post-merger will vary depending on the model, which we analyse below. 
\begin{center}
\begin{table}
{\small
\hfill{}
\begin{tabular}{l c c}
    \hline
        & \multicolumn{2}{c}{Time (Gyr)} \\
\cline{2-3}
        & Fiducial Resolution &  Low Resolution \\
\hline
\hline
Model SDH  &    1.16 &    1.31 \\
Model BS   &     --- &    ---  \\
Model ONB  &    1.00 &    1.00 \\
Model WT   &    1.02 &    1.00 \\ 
Model DQM  &    0.98 &    0.98 \\ 
\hline
Model BSw  &    1.13 &    1.10 \\ 
Model ONBc &    1.10 &    1.04 \\ 
Model WTh  &    1.02 &    1.01 \\ 
Model DQMe &    0.87 &    0.86 \\
\hline
\end{tabular}}
\hfill{}
\caption{Merger time of the black holes, as measured from the beginning of the simulation.  Dashed lines indicate that the black holes did not merged within our 1.5 Gyr simulation time.}
\label{r_mergertime} 
\end{table}
\end{center}

Consider two identical black holes in the merged core during the black hole merger.  First, the total region of influence of the unmerged black holes will be between $V = \frac{4}{3}\pi r_\text{inf}^3$ and $2V$ depending on separation of the black holes.  Since $r_\text{inf}$ is independent of black hole mass, the total region of influence after the merger will simply be $V$.  In all cases, $\dot{E} \propto \dot{M}$, so the amount of feedback energy available prior to the merger will be proportional to $\dot{M}_\text{BH$_1$} + \dot{M}_\text{BH$_2$}$; after the merger, we will have energy proportional to $\dot{M}_\text{BH$_1$+BH$_2$}$; if we define $\dot{M}$ to be the accretion rate of one black hole prior to the merger, then for each accretion rate we studied, across the merger we have
\begin{align*}
2\dot{M}_\text{Edd} & \rightarrow 2\dot{M}_\text{Edd} ,\\
2\dot{M}_\text{B}   & \rightarrow 4\dot{M}_\text{B}   ,\\
2\dot{M}_\text{drag}& \rightarrow  \dot{M}_\text{drag},\\ 
2\dot{M}_\text{visc}& \rightarrow  \dot{M}_\text{visc}.
\end{align*}
We can clearly see that the only total accretion rate in the core that remains unchanged by the merger is $\dot{M}_\text{Edd}$; the total Bondi accretion rate doubles and the total viscous and drag accretion rates halves.  Thus, in Models SDH, BS and WT (which use Bondi accretion), there is more feedback energy available after the merger than before, and this increased amount of energy will be distributed to a smaller region.  Likewise, there is less feedback energy available in Models ONB and DQM after the merger, which will also be distributed within a smaller volume.  Thus, we can argue that the galactic core becomes more energetic assuming Bondi accretion, but less energetic assuming viscous or drag accretion when two black holes merge.

In the models, Model DQM has the least stringent merger prescription, with only a separation criteria.  As expected, this is the first model to merge.  The most stringent merging criteria is in Model BS, which is never met.  The circular velocity at the more massive black hole's smoothing length is typically ten times smaller than the local sound speed, and the relative velocity between the black holes never drops below this threshold.  Moreover, the chaotic motion and two-body interactions of the black holes typically keeps them further apart than the separation criteria.  

\subsection{Global star formation rates}
\label{r_gsfr}

\begin{figure}
\begin{center}
\includegraphics[width=1.0\columnwidth]{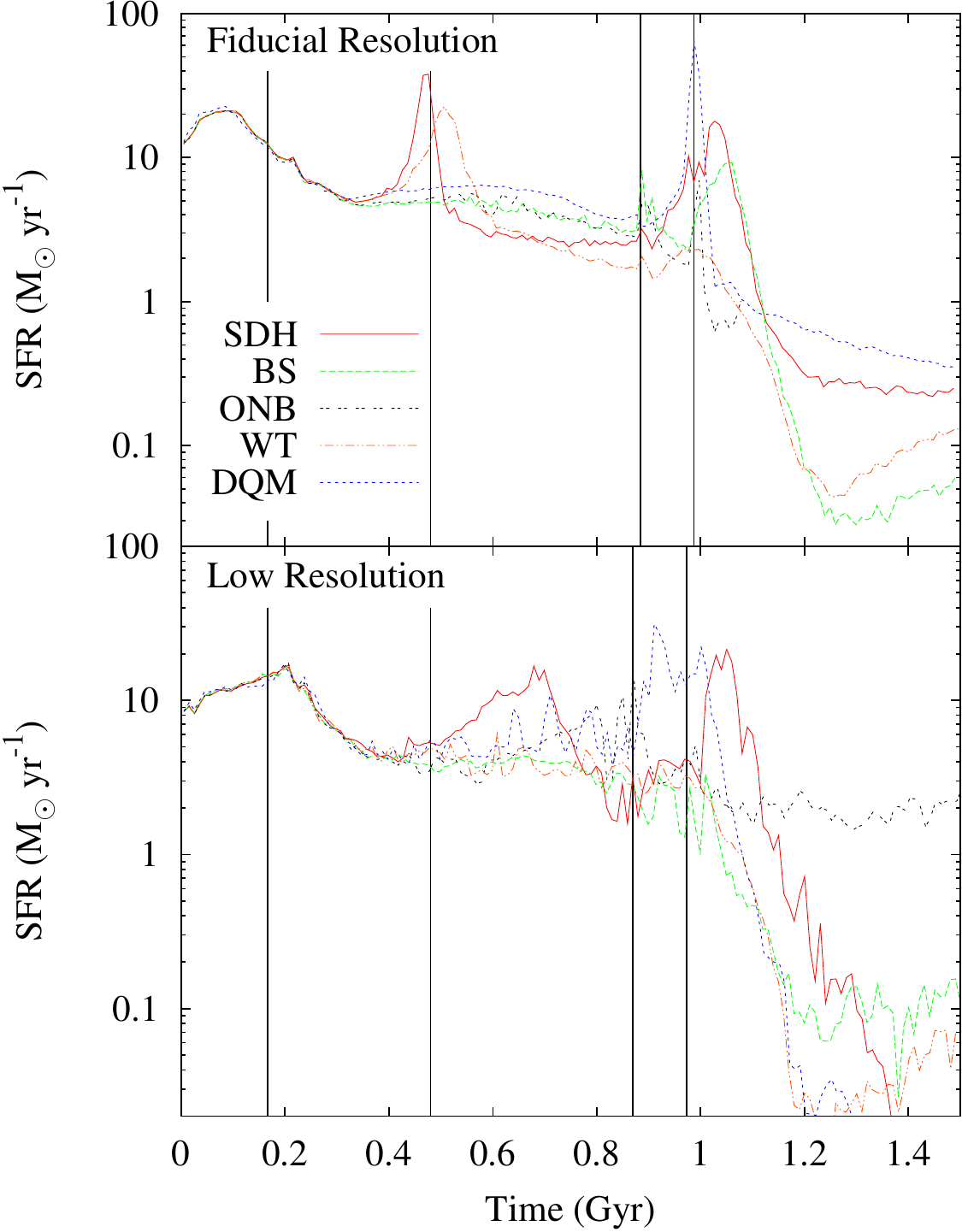}
\end{center}
\caption{Global star formation rate for the fiducial (top) and low resolution (bottom) runs.  The vertical lines represent the times of first periapsis, apoapsis, second periapsis and core merger for the dynamic black hole mass models, as given in Table \ref{r_imptime}.}\label{gsfr} 
\end{figure}

During a major merger, there is typically an burst of star formation at apoapsis and again during core merger due to the infall of gas on to the galactic core (e.g. \citet{MH96}; SDH05; DQM11).  There are several characteristics that can modify the star formation rate at these epochs, including initial conditions and the inclusion of a galactic bulge and/or AGN feedback (SDH05).  In {\sc Hydra}, the star formation algorithm differs from those used in SDH05, BS09, ONB08 and DQM11, so we do not expect an exact reproduction of their results; however, any differences in SFRs amongst our models will be solely the result of the AGN feedback algorithm and its interaction with the star formation algorithm.  The time-averaged SFR for the fiducial (top panel) and the low resolution (bottom panel) are plotted in Fig. \ref{gsfr}.  

In these models, there are notable star formation bursts shortly after the beginning of the simulation, at apoapsis and again at core merger.  In all cases, the initial burst is a result of the gas disc relaxing \citep{WT12}; this initial burst also exists in simulations of an isolated galaxy that is constructed in a similar manner to the one presented here.  SDH05 and DQM11 create the vertical structure of their gas disc by assuming hydrostatic equilibrium after setting their equation of state.  Although cooling is not considered during this process, it means that their initial conditions and simulations are more equivalent than ours.  Thus, as expected, this initial burst of star formation is missing from their results.  This initial burst is resolution dependent, being higher and shorter lived in the fiducial resolution models.  After this epoch in the low resolution models, the only outstanding features are post-apoapsis burst in Model SDH$_\text{l}$, the core merger burst for Models SDH$_\text{l}$ and DQM$_\text{l}$, and the expected drop in SFRs for all models (except Model ONB$_\text{l}$) shortly after core merger.  The low resolution features do not necessarily have counterparts in the fiducial resolution simulations and vice versa, as a result of the stellar feedback being resolution dependent:  low resolution star formation events extend over a considerably larger volume and mass than fiducial resolution events due to the larger gas particle mass and lower spatial resolution.  Thus, a low resolution event has an unavoidably greater impact on its environment, and causes comparatively large regions of the galaxy to undergo a cooling delay.  This explains the delayed star formation peak after apoapsis in Model SDH$_\text{l}$, the lack of peak in Model WT$_\text{l}$ after apoapsis, and generally lower SFRs at core merger.

In Model SDH, the SFR increases by 0.9 dex at both apoapsis and core merger; in SDH05, there is no burst at apoapsis and there is an increase of 1 dex at core merger.  This burst at core merger exists in all the models, except Model WT, where the high feedback rate of thermal energy prevents the gas from cooling in to the regime where star formation can commence.  After the core merger burst, the SFRs drop, as in SDH05, due to the heating and total disruption of the system.  

In Model WTh, the star formation burst at apoapsis peaks approximately 50 Myr earlier than in Model WT.  The apoapsis burst is a result of a nuclear inflow of gas, and the lower value of $h_\text{min}$ allows higher densities to be resolved sooner; thus inflowing gas enters the permissible density regime of star formation for Model WTh before Model WT.  For the same resolution argument, Model WTh also has a small burst of star formation (an increase of 0.6 dex) at core merger.  This difference in the SFR highlights one of the well known drawbacks of resolution dependency in the star formation algorithm, although developing a model that is independent of resolution awaits a full understanding of star formation in a global context.

The SFR in Model ONB remains relatively smooth until a small increase of 0.6 dex at core merger.  Even after core merger, the SFR remains relatively unchanged due high core densities and moderate core temperatures.

In DQM11, there is a major star formation burst shortly after first periapsis and again starting at second periapsis (see their fig. 1, middle panel); Model DQM reproduced the burst at second periapsis.  The hot ring around the void means that the star formation must occur in the galactic disc, which results in the lack of peak after first periapsis.  At second periapsis, the merger is efficient enough to create a galactic core, causing the core merger star formation burst.  After this final burst, the SFR drops to small values, agreeing with DQM11.

\section{Final States}
\label{fstates}

We evolved our galaxy merger simulations for 1.5 Gyr, and the characteristics of the remnants are presented here.  The exception is the fiducial resolution Model ONB, which was evolved for 1.1 Gyr.  From our analysis of Model ONB$_\text{l}$, we do not expect any significant evolution between 1.1 and 1.5 Gyr.  Therefore, for the final remnant of Model ONB, we present a combination of data from 1.1 Gyr, data extrapolated to 1.5 Gyr, and data from Model ONB$_\text{l}$; we will clearly state where the data comes from in each instance.

\begin{figure}
\begin{center}
\includegraphics[width=1.0\columnwidth]{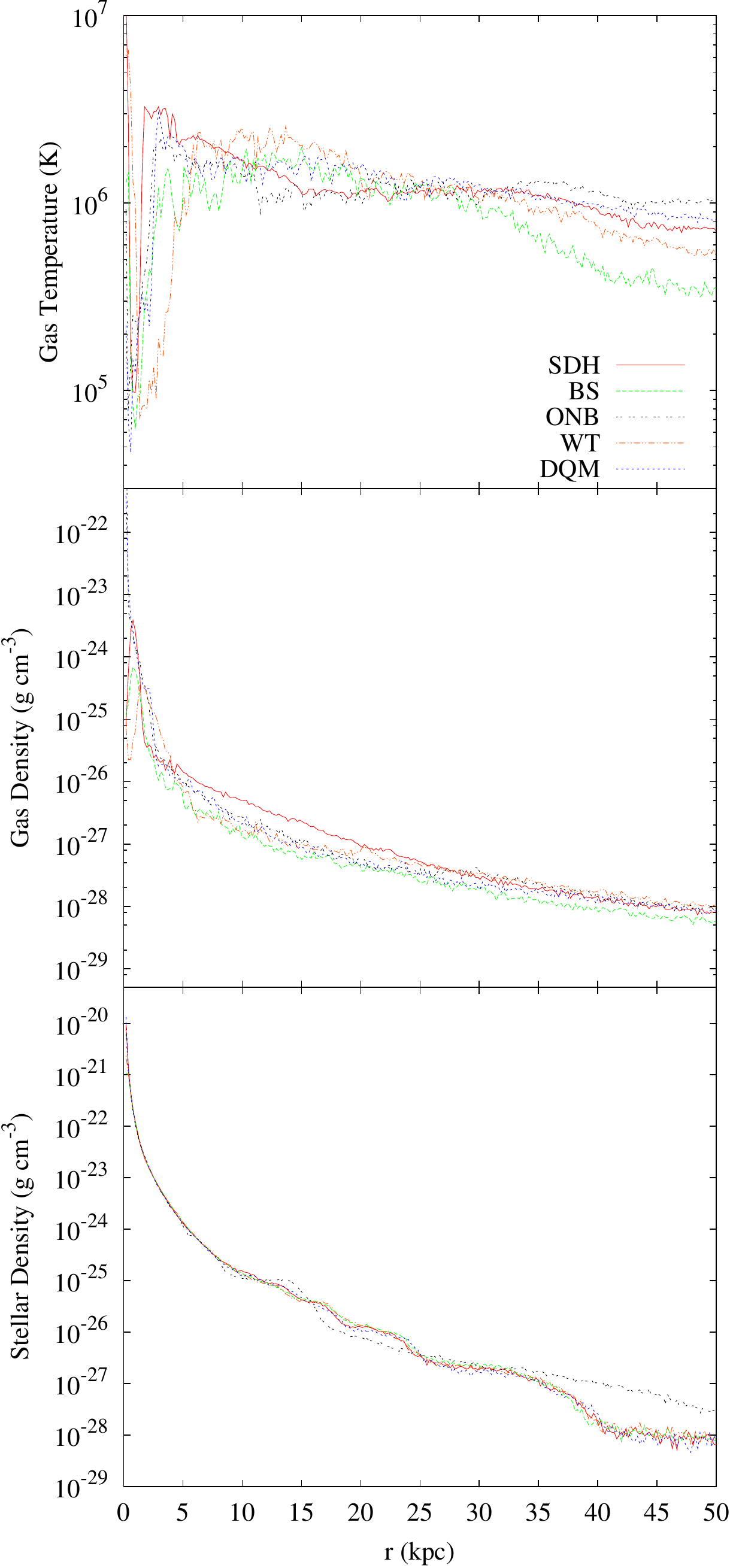}
\end{center}
\caption{Radial profiles of our remnants, averaged over all azimuthal and polar angles.  From top to bottom, we show gas temperature, gas density, and stellar density. The profile for Model ONB is taken at 1.1 Gyr.}\label{f_profiles} 
\end{figure}

\begin{figure*}
\begin{center}
\includegraphics[width=0.24\textwidth]{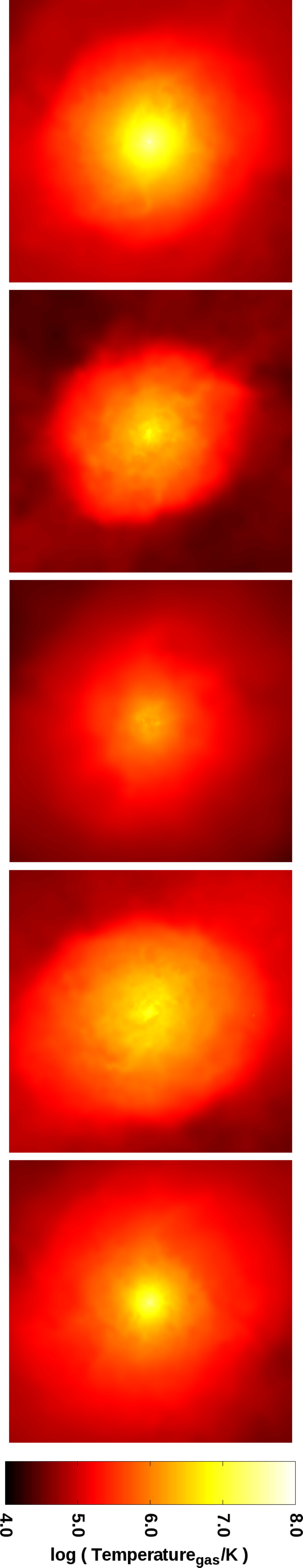}
\includegraphics[width=0.24\textwidth]{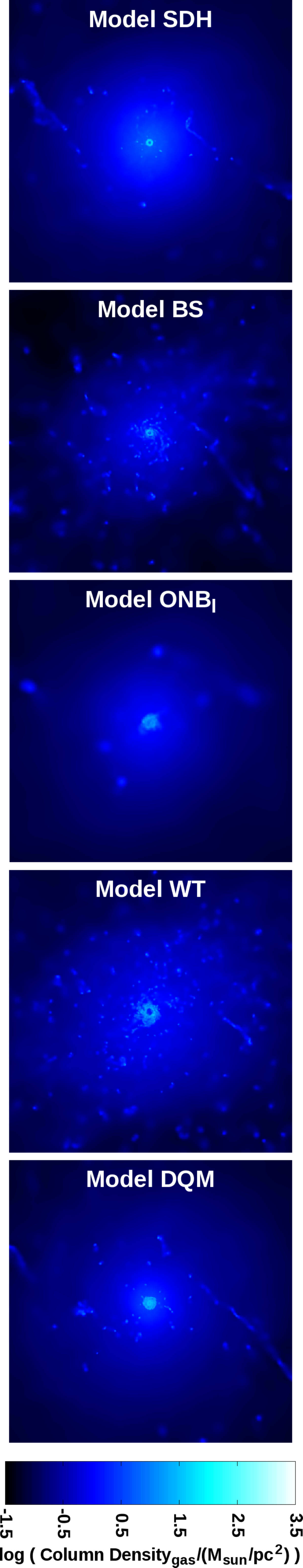}
\includegraphics[width=0.24\textwidth]{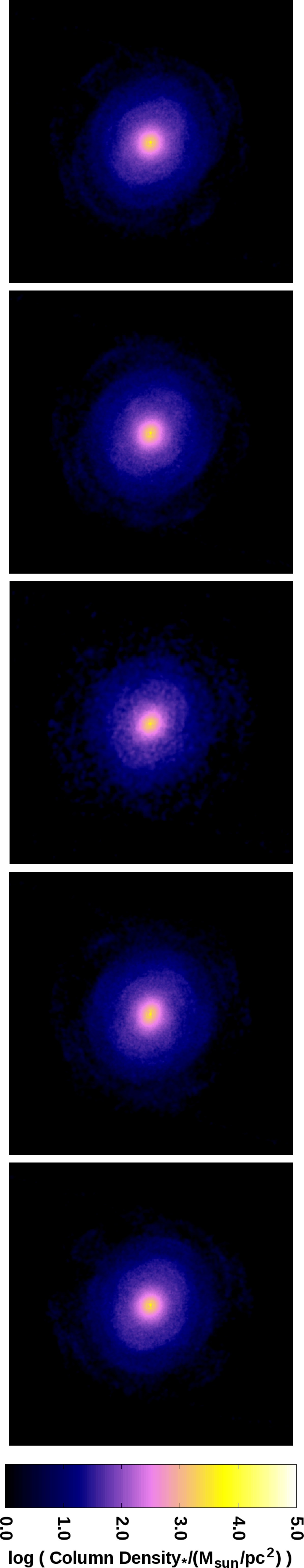}
\end{center}
\caption{ \emph{Left to right}: Face-on gas temperature, gas column density, and stellar column density of each remnant.  Each frame measures 100 kpc per side, with a image resolution of 195 pc pixel$^{-1}$ (391 pc pixel$^{-1}$) for the fiducial (low) resolution models.}\label{f_final5}  
\end{figure*}

\begin{figure*}
\begin{center}
\includegraphics[width=0.33\textwidth]{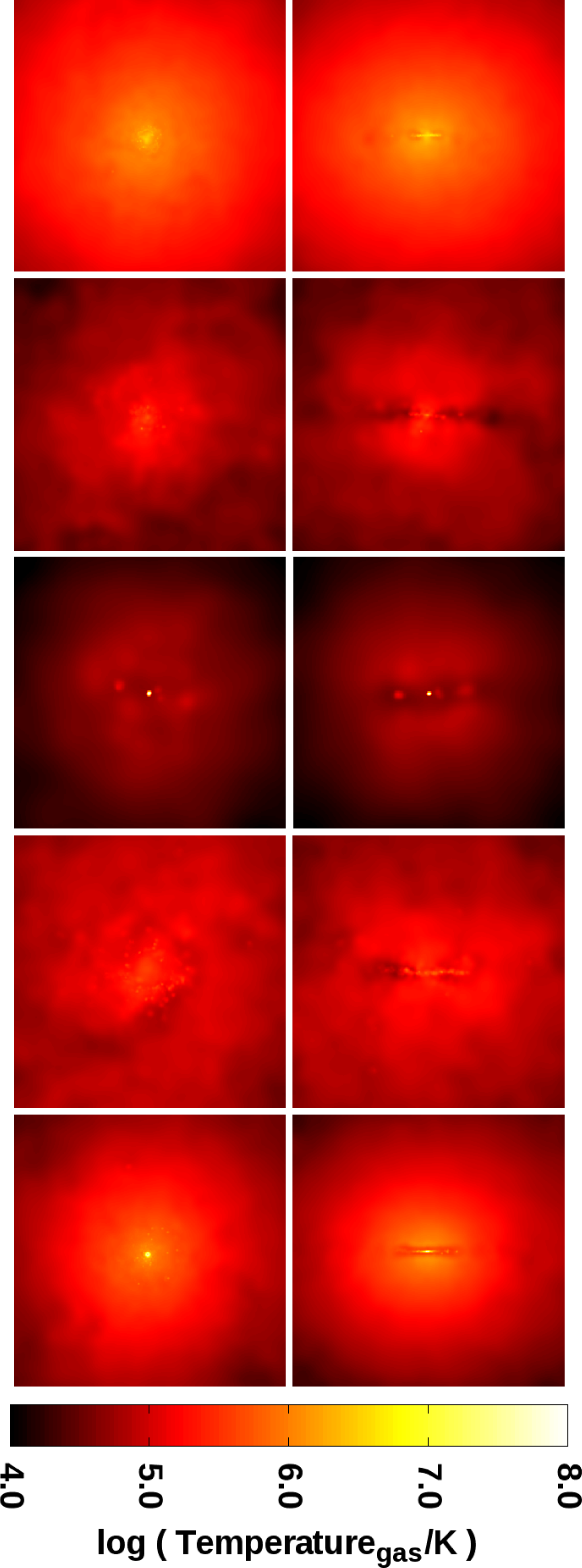}
\includegraphics[width=0.33\textwidth]{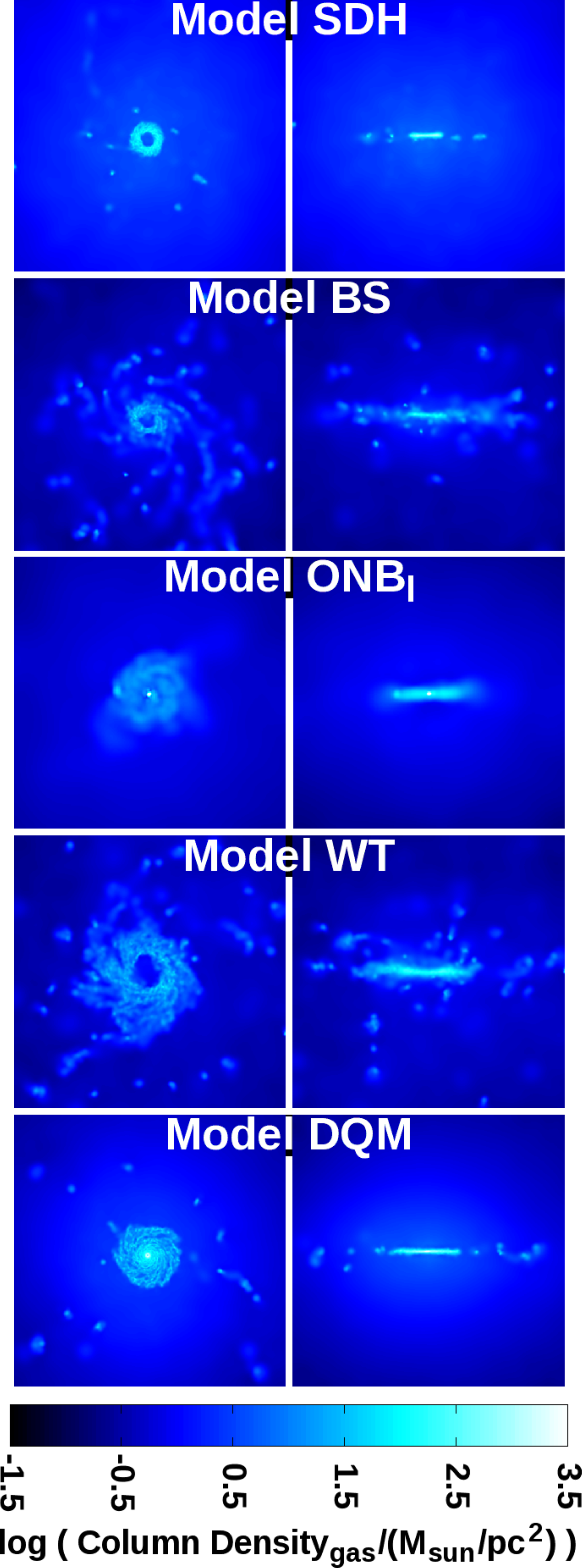}
\includegraphics[width=0.33\textwidth]{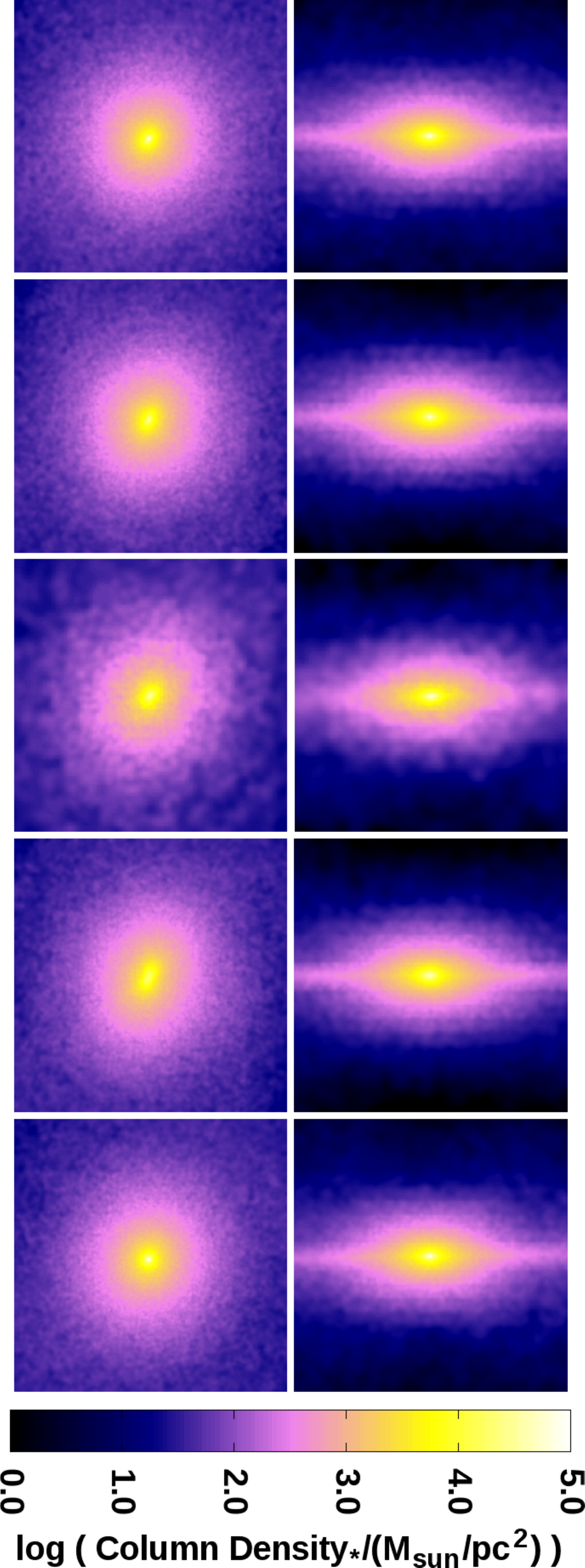}
\end{center}
\caption{ \emph{Left to right}: Face-on and edge-on gas temperature, face-on and edge-on gas column density, and face-on and edge-on stellar column density of each remnant.  Each frame measures 20 kpc per side, with a image resolution of 39 pc pixel$^{-1}$ (78 pc pixel$^{-1}$) for the fiducial (low) resolution models.}\label{f_final10}  
\end{figure*}

The final remnant in all models is a reformed gas and stellar disc surrounded by a fragmented hot gas halo.  The radial profiles for gas temperature, gas density and stellar column density (averaged over all azimuthal and polar angles) for each remnant are similar, as seen in Fig. \ref{f_profiles}; the values typically span 0.5 dex at any given radius.  
In Figs. \ref{f_final5} and \ref{f_final10}, we have plotted the final gas temperature, gas column density and stellar column density for frames of 100 and 20 kpc per side, respectively. 
 
\subsection{Gas properties}
By the end of the simulation, the gas that was originally in a galactic disc is heavily depleted while the majority of the gas that was originally in the halo remains; the fraction of remaining gas for each fiducial resolution model is given in Table \ref{f_finalstargas}.  The majority of the depleted gas has been converted in to stars as opposed to being accreted on to a black hole.  The gas that is ejected from the disc has temperatures and pressures a few orders of magnitude greater than the halo gas, thus it can easily escape from the system.  Therefore, the gas halo does not play a roll in recycling the ejected gas in to stars. 
\begin{center}
\begin{table*}
{\small
\hfill{}
\begin{tabular}{l r r r r r r}
\hline
Model & $\frac{M_\text{stars,f}}{M_\text{stars,i}}$ & $\frac{M_\text{gas,f}}{M_\text{gas,i}}$ & $\frac{M_\text{disc gas,f}}{M_\text{disc gas,i}}$ & $\frac{M_\text{halo gas,f}}{M_\text{halo gas,i}}$ & $\frac{M_\text{BH,f}}{M_\text{BH,i}}$ & $\frac{M_\text{acc}}{\Delta M_\text{gas}}$ \\ 
\hline
\hline
SDH  & 1.089 & 0.615 & 0.205 & 0.988 & 124.7 & 0.00286 \\ 
BS   & 1.056 & 0.756 & 0.497 & 0.992 &  27.9 & 0.00096 \\ 
ONB  & 1.053 & 0.772 & 0.526 & 0.996 &  40.6 & 0.00165 \\ 
WT   & 1.061 & 0.734 & 0.452 & 0.992 & 137.6 & 0.00450 \\ 
DQM  & 1.098 & 0.577 & 0.127 & 0.987 & 165.9 & 0.00339 \\
\hline
BSw  & 1.046 & 0.802 & 0.592 & 0.992 &  16.6 & 0.00069 \\ 
ONBc & 1.055 & 0.766 & 0.513 & 0.996 &   1.4 & 0.00000 \\ 
WTh  & 1.065 & 0.716 & 0.415 & 0.991 & 169.8 & 0.00521 \\ 
DQMe & 1.064 & 0.720 & 0.423 & 0.991 & 615.1 & 0.01917 \\
\hline
\end{tabular}}
\hfill{}
\caption{Ratios of final to initial global mass components for each of our fiducial resolution models; black hole masses are dynamic masses (except for Model DQM) since gas is removed in discrete amounts of $m_\text{gas}$.  Gas labelled as disc gas or halo gas is based upon its initial position.  We define $M_\text{acc}$ as the total mass of gas (dynamically) accreted by black holes and $\Delta M_\text{gas} = M_\text{gas,i} - M_\text{gas,f}$.  Depleted gas that is not accreted on to a black hole has been converted into stars.  The values for Models ONB and ONBc are extrapolated to 1.5 Gyr based upon data from the final output and their low resolution counterparts.  Specifically, we assume an average SFR of 1.0 M$_{\astrosun}$ yr$^{-1}$ (1.6 M$_{\astrosun}$ yr$^{-1}$) and $\dot{M}_{BH} = 5.6\times 10^{-3} $M$_{\astrosun}$ yr$^{-1}$ ($6.0\times 10^{-8} $M$_{\astrosun}$ yr$^{-1}$) for the remainder of the simulation for Model ONB (ONBc).}
\label{f_finalstargas} 
\end{table*}
\end{center}

\begin{figure}
\begin{center}
\includegraphics[width=1.0\columnwidth]{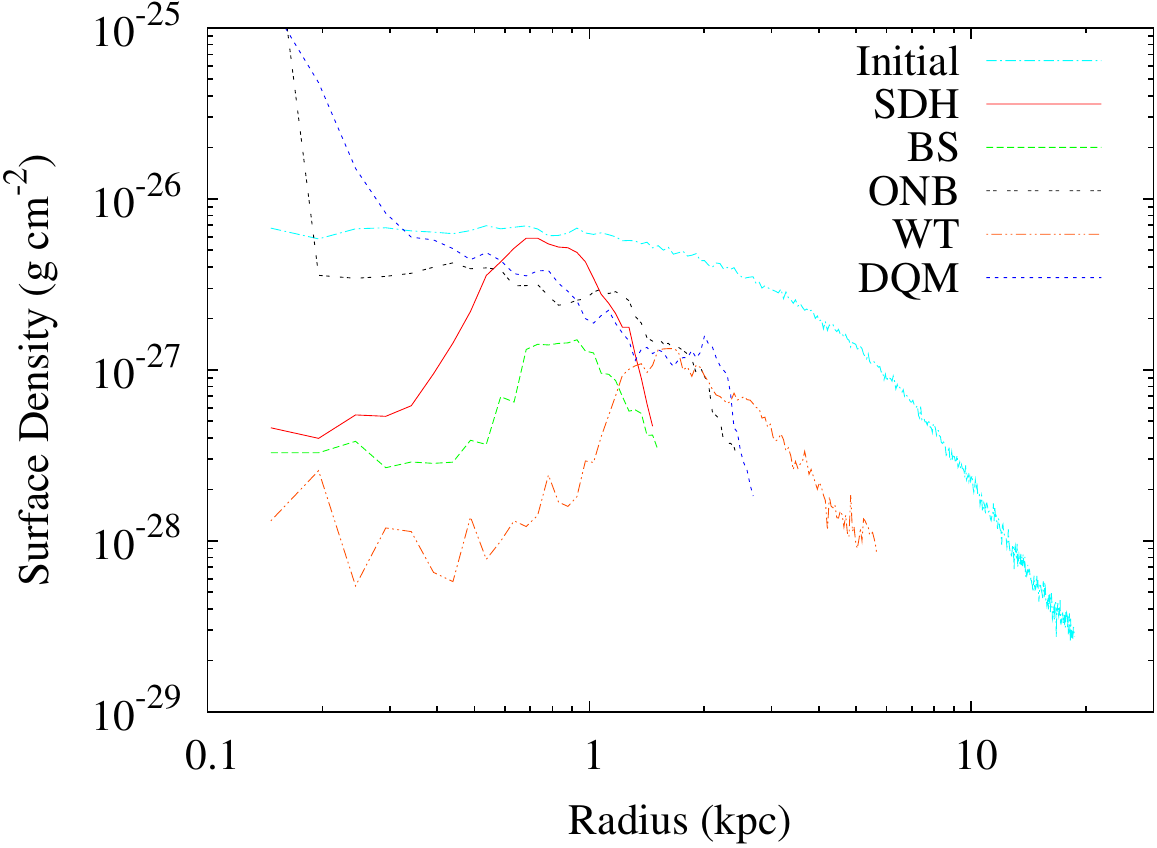}
\end{center}
\caption{Gas surface density profile, averaged over all azimuthal angles, for the initial and final discs of the fiducial resolution models. The profiles are truncated at the edge of the disc, and plotted in bins of 49 pc.  The remnant profile for Model ONB is taken at 1.1 Gyr.}\label{f_finalsurfden}  
\end{figure}

The amount of substructure varies considerably amongst remnants.  Model ONB's remnant is a well formed disc in a smooth halo, while Models BS and WT have a loosely reformed disc in a fragmented halo.  Shortly after core merger, Model BS undergoes an outburst event which expels most of the gas, and by 1.5 Gyr, this gas is starting to recondense on to the core.  In Model WT, the remaining gas cools and fragments, and then begins to recondense on to the core.  Thus, in both models there are similar remnants at 1.5 Gyr but different histories prior to this time.  In all cases, a small disc begins to reform; the surface density profile for the reformed fiducial resolution discs is plotted in Fig. \ref{f_finalsurfden}.
The scale lengths of the discs embedded within the stellar remnant are approximately 0.5 kpc, compared to the initial value of 2.46 kpc.  There are voids in the centre of the discs in Models SDH, BS and WT, thus the peak surface density occurs at 0.7--1.5 kpc from the centre.  The voids persist from the residual angular moment of the gas and from the weak (but remaining) AGN feedback.

\subsection{Stellar remnant}
\begin{figure*}
\begin{center}
\includegraphics[width=1.0\textwidth]{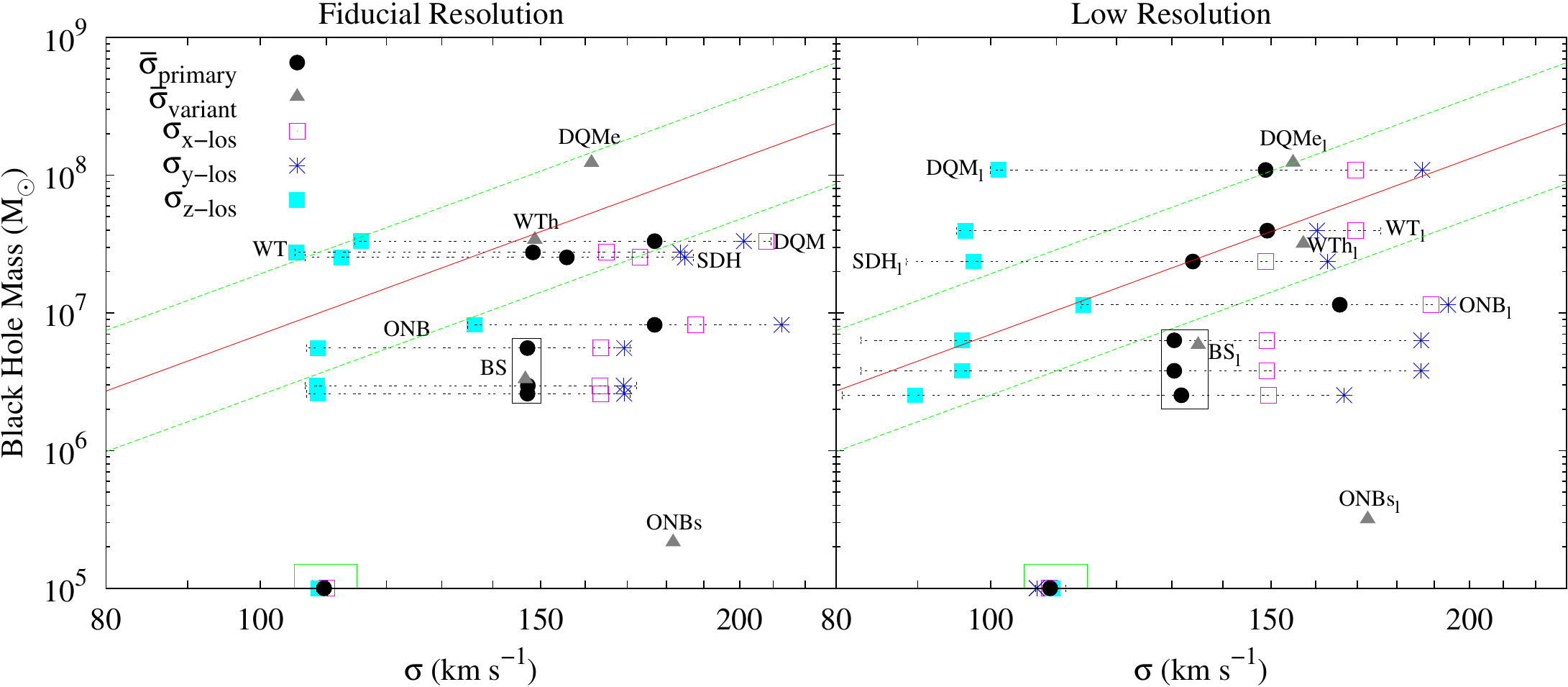}
\end{center}
\caption{The $M_\text{BH}$--$\sigma$ relation for our simulations, along with the observed relation (red) and the one-sigma scatter (dashed) from \citet{Getal09}.  For our five primary models at each resolution, the dot represents the average $\sigma$ of 1000 random lines of sight, the horizontal bars represent the range of all calculated $\sigma$'s, and the remaining three symbols on the horizontal line represent $\sigma$ taken along the $\pm x-$, $\pm y-$ and $\pm z-$lines of sight.  For our variant models, we have only plotted the average $\sigma$ of  1000 random lines of sight (solid triangles).  We have explicitly labelled which points/set of points corresponds to which models.  The black dots in the red rectangles represent the black holes in Model BS; the lower two points represent the actual black holes, and the upper point is calculated assuming one black hole with $M_\text{BH} = M_\text{BH$_1$} + M_\text{BH$_2$}$ at the centre of mass of the black hole system.  The solid triangle in the red rectangle is the result for Model BSw.  The points in the green box are our initial relationships.  The data for Models ONB and ONBc are taken at 1.1 and 1.25 Gyr, respectively.}\label{msigma} 
\end{figure*}

For each remnant, we calculate the stellar velocity dispersion, $\sigma$, around the black hole.  To mimic observations, we calculate this using
\begin{equation}
\label{msigmaObvs}
\sigma^2 = \frac{\int_0^{R_\text{e}} \! \sigma^2_\text{los}(r) I(r) r \, \mathrm{d} r}{\int_0^{R_\text{e}} \! I(r) r \, \mathrm{d} r},
\end{equation}
where $\sigma_\text{los}$ is the line of sight velocity dispersion, $I(r)$ is the projected 2D stellar density profile, and $R_\text{e}$ is the half-light (mass) radius; this is similar to what is done numerically in DQM11 and observationally in \citet{Getal09}.  Since the final stellar configuration is highly triaxial (see the two right-most columns in Fig. \ref{f_final10}), the $\sigma$ we present in Fig. \ref{msigma} is averaged over 1000 random lines of sight; we have also included the full range of $\sigma$'s calculated, as well as the preferentially chosen lines of sight along the $\pm x-$, $\pm y-$ and $\pm z-$directions.

For the fiducial (low) resolution models, the velocity dispersion spans a range of $\sim$$75\pm15$ km s$^{-1}$ ($\sim$$90\pm15$ km s$^{-1}$), and is latitude dependent.  Since the remnant reforms a disc, there is a large $\sigma$ if the line of sight lies in the plane of the disc, and $\sigma$ decreases as the elevation increases.  Except for Models BS and ONB, the average $\sigma$ lies is within the one-sigma scatter of the observational relation.  

In Model BS, the black holes never merge and do not fall within one-sigma of the observed relation; the black holes in Model BSw merge, but still do not fall within one-sigma of the observed relation.  This indicates that the difference from observational expectations is not a result of the lack of merger for the model parameters we have implemented.  In these models, between apoapsis and second periapsis, the accretion rates are lower than the other models, which hinders black hole growth.  Also, randomly depositing feedback energy around the black hole, rather than returning it isotropically or kernel-weighted, could increase the velocity dispersion of the stars by increasing the anisotropy of the gas.

In Model ONB, the black hole mass is 9.5 times lower than predicted by the $M_\text{BH}$--$\sigma$ relation, assuming the stellar velocity dispersion remains constant; a mass 3.4 times larger would bring the point within the one-sigma scatter of the relation.  Since the feedback energy is returned far from the galactic core, gravity is the only influence the black hole has on the nearby stars.  Thus, the self-regulating feedback mechanism is never fully realised, which could explain the lack of agreement with the observed relation.  This is also true for Model ONBc, but to a more extreme extent since the low accretion rate means that the black hole only negligibly grows and cannot couple to the stellar system.  

\section{Conclusions}
\label{conclusion}

This paper details a study of five different AGN feedback algorithms, which were all run using the numerical code {\sc Hydra}, started from the same initial conditions, and implemented the same star formation algorithm; two resolutions were tested for each model.  The AGN feedback algorithms used in Models SDH, BS, ONB and DQM were previously found in the literature and the algorithm used in Model WT was developed specifically for this study.  We also ran four additional models, each of which was a slight variation of one of the primary models to study a specific aspect of the model.

We have performed a full analysis of all models, but only presented fiducial resolution results of the five primary models and selected other results.  Although the low resolution models run considerably faster, their results are noisy, and some of the physics is essentially lost.  However, the low resolution studies are a good test of the varying feedback models, and provide a first-order comparison between the models.  Furthermore, the continued implementation of cosmological simulations mean these resolutions will inevitably be used in the future.  

By comparing these models using the same initial conditions, star formation algorithm and numerical code, any differences obtained are a direct result of the AGN feedback.  Although we have tried to isolate each component of the AGN feedback algorithm, we accept that they are all intimately intertwined, and the effects cannot necessarily be disentangled from one another.  We also accept that every simulation we run involves free parameters; although we have -- as best as possible -- matched them to their source simulations from the literature, it is possible that we can modify these parameters such that all simulations will return similar final states.  Since the goal of this paper was to compare different feedback models and not to analyse any given model in great detail, we defer to the source literature for discussions on the sensitivity of the results to the free parameters.  We thus refrain from making strong statements about the `accuracy' of given models.  Our principal conclusions of the five key components are as follows:

\begin{enumerate}

\item The black hole advection algorithm plays a key role since small displacements can cause great changes in the accretion rate.  Coupling the black hole to a gas particle can allow for oscillations or chaotic motion if a void is formed around the black hole, or if the nearby gas does not meet the velocity criteria.  Using a tracer mass yields smooth movement of the black hole, but the evolution time of the merger is decreased and disc morphology is altered in non-trivial ways such as preventing the formation of a bar. This obviously impacts other physical properties, most notably the SFR as well. Clearly evolving the mass of the black hole in a way that does not overly impact the expected evolution of the disc is an important goal. The algorithms that limit the distance a black hole can be displaced per iteration and where the direction is based upon the local stellar or total mass distribution appear to be optimal.

\item We test four different accretion rates: $\dot{M}_\text{B}(\alpha=100)$, $\dot{M}_\text{B}(\alpha \equiv \alpha(n_\text{H}))$, $\dot{M}_\text{drag}$ and $\dot{M}_\text{visc}$.  Each model yields a different accretion history, with substantially different qualitative profiles and quantitative differences of up to a factor of three orders of magnitude at any given time.  The total black hole mass in the remnants varies by factor of 6.9, with Model BS being the least massive, and DQM being the most massive.  The three models that fall within one-sigma scatter of the $M_\text{BH}$--$\sigma$ relation have final black hole masses that differ by less than 30 per cent.

\item The form of feedback can drastically modify the large- and small-scale systems.  Delivering the feedback to the halo gas leads to little modification of the core region, resulting in high core densities and nominal temperatures.  Thermal feedback delivered to the core region can drive strong outflows.  Kinetic feedback delivered to the core region efficiently creates a void around the black hole, resulting in low gas temperature and density.  This feedback is persistent and efficient, keeping the core properties nearly constant for all time; only a cataclysmic event, such as a core merger, can modify these characteristics.  

\item We tested three categories of particle accretion algorithms: stochastic-unconditional, stochastic-conditional and continual-conditional; two of the three continual-conditional algorithms randomly selected \emph{which} particles were accreted but never \emph{how many} or \emph{how often}.  The continual-conditional cases gives the best agreement between the dynamical and internal mass; here, the discrepancy is never more than 2$m_\text{g}$.  The stochastic-unconditional algorithm of Model SDH, however, contains discrepancies up to 15$m_\text{g}$.  Thus, for agreement between dynamical and internal masses, continual-conditional algorithms appear optimal.

\item In Models BS and BS$_\text{l}$, the black holes never merge; in these models, there is considerable chaotic movement of the black holes in the remnant core, preventing both merger criteria from being simultaneously met.  In the remainder of the models, the black holes merge during or shortly after core merger, as one would reasonably expect.  The importance of a `reasonable' merger time is that the amount of feedback energy available increases across the merger for Bondi accretion and decreases for drag and viscous accretion.

\end{enumerate}

The models presented here do not represent an exhaustive list.  For example, an accretion rate suggested by \citet{HPNK12} is an interpolation between Bondi accretion and a free-fall accretion rate.  \citet{PNK11} suggest using an accretion disc particle; in this two-tier accretion process, gas is first accreted on to an accretion disc, and then the gas is accreted from the disc on to the black hole using a viscous time-scale.  The accretion disc particle method of accretion has recently been implemented in major merger simulations similar to those presented here \citep{WT13power}.  

Lastly, while it is possible that free parameters in different models can be adjusted such that all yield similar final states, it is very clear that this will not produce agreement throughout the evolution. While we have a vast
amount of observational data of many stages of major mergers, including final remnants, it is difficult to build these various snapshots into an obvious picture of the evolutionary processes, and the fundamental theoretical development of AGN modelling still has much to contribute in this regard. As we have shown here, given the many different models of AGN feedback, building an understanding of the evolution of mergers is an important and challenging task.

\section*{Acknowledgments}

We thank the anonymous referee for helpful comments that improved the clarity and overall quality of the manuscript. We thank Larry Widrow for his NFW halo generator.  We also thank David Williamson for useful discussions.  JW is supported by NSERC and Saint Mary's University.  RJT is supported by a Discovery Grant from NSERC, the Canada Foundation for Innovation, the Nova Scotia Research and Innovation Trust and the Canada Research Chairs Program. Simulations were run on the CFI-NSRIT funded {\em St Mary's Computational Astrophysics Laboratory}.

\bibliography{WTbib}

\begin{thebibliography}{99}
\expandafter\ifx\csname natexlab\endcsname\relax\def\natexlab#1{#1}\fi

\bibitem[{{Berentzen} {et~al}\mbox{.}(2009){Berentzen}, {Preto}, {Berczik},
  {Merritt}, \& {Spurzem}}]{BPBMS09}
{Berentzen} I., {Preto} M., {Berczik} P., {Merritt} D., {Spurzem} R., 2009,
  \apj, 695, 455

\bibitem[{{B{\^i}rzan} {et~al}\mbox{.}(2004){B{\^i}rzan}, {Rafferty},
  {McNamara}, {Wise}, \& {Nulsen}}]{BRMWN04}
{B{\^i}rzan} L., {Rafferty} D.~A., {McNamara} B.~R., {Wise} M.~W., {Nulsen}
  P.~E.~J., 2004, \apj, 607, 800

\bibitem[{{Bode} {et~al}\mbox{.}(2012){Bode}, {Bogdanovi{\'c}}, {Haas},
  {Healy}, {Laguna}, \& {Shoemaker}}]{BBHHLS12}
{Bode} T., {Bogdanovi{\'c}} T., {Haas} R., {Healy} J., {Laguna} P., {Shoemaker}
  D., 2012, \apj, 744, 45

\bibitem[{{Boehringer} {et~al}\mbox{.}(1993){Boehringer}, {Voges}, {Fabian},
  {Edge}, \& {Neumann}}]{BVFEN93}
{Boehringer} H., {Voges} W., {Fabian} A.~C., {Edge} A.~C., {Neumann} D.~M.,
  1993, \mnras, 264, L25

\bibitem[{{Bondi}(1952)}]{B52}
{Bondi} H., 1952, \mnras, 112, 195

\bibitem[{{Booth} \& {Schaye}(2009)}]{BS09}
{Booth} C.~M., {Schaye} J., 2009, \mnras, 398, 53, \ (BS09)

\bibitem[{{Brook} {et~al}\mbox{.}(2004){Brook}, {Kawata}, {Gibson}, \&
  {Freeman}}]{BKGF04}
{Brook} C.~B., {Kawata} D., {Gibson} B.~K., {Freeman} K.~C., 2004, \apj, 612,
  894

\bibitem[{{Buff} \& {McCray}(1974)}]{BM74}
{Buff} J., {McCray} R., 1974, \apj, 189, 147

\bibitem[{{Cabrit} {et~al}\mbox{.}(1990){Cabrit}, {Edwards}, {Strom}, \&
  {Strom}}]{CESS90}
{Cabrit} S., {Edwards} S., {Strom} S.~E., {Strom} K.~M., 1990, \apj, 354, 687

\bibitem[{{Castor} {et~al}\mbox{.}(1975){Castor}, {Abbott}, \& {Klein}}]{CAK75}
{Castor} J.~I., {Abbott} D.~C., {Klein} R.~I., 1975, \apj, 195, 157

\bibitem[{{Cavaliere} \& {Fusco-Femiano}(1976)}]{CF76}
{Cavaliere} A., {Fusco-Femiano} R., 1976, \aap, 49, 137

\bibitem[{{Churazov} {et~al}\mbox{.}(2002){Churazov}, {Sunyaev}, {Forman}, \&
  {B{\"o}hringer}}]{CSFB02}
{Churazov} E., {Sunyaev} R., {Forman} W., {B{\"o}hringer} H., 2002, \mnras,
  332, 729

\bibitem[{{Ciotti} \& {Ostriker}(2001)}]{CO01}
{Ciotti} L., {Ostriker} J.~P., 2001, \apj, 551, 131

\bibitem[{{Couchman}(1991)}]{C91}
{Couchman} H.~M.~P., 1991, \apjl, 368, L23

\bibitem[{{Couchman} {et~al}\mbox{.}(1995){Couchman}, {Thomas}, \&
  {Pearce}}]{CTP95}
{Couchman} H.~M.~P., {Thomas} P.~A., {Pearce} F.~R., 1995, \apj, 452, 797

\bibitem[{{Cowie} {et~al}\mbox{.}(1978){Cowie}, {Ostriker}, \& {Stark}}]{COS78}
{Cowie} L.~L., {Ostriker} J.~P., {Stark} A.~A., 1978, \apj, 226, 1041

\bibitem[{{Cowie} {et~al}\mbox{.}(1996){Cowie}, {Songaila}, {Hu}, \&
  {Cohen}}]{CSHC96}
{Cowie} L.~L., {Songaila} A., {Hu} E.~M., {Cohen} J.~G., 1996, \aj, 112, 839

\bibitem[{{Croft} {et~al}\mbox{.}(2009){Croft}, {Di Matteo}, {Springel}, \&
  {Hernquist}}]{CDSH09}
{Croft} R.~A.~C., {Di Matteo} T., {Springel} V., {Hernquist} L., 2009, \mnras,
  400, 43

\bibitem[{{Crowther}(2007)}]{C07}
{Crowther} P.~A., 2007, \araa, 45, 177

\bibitem[{{Debuhr} {et~al}\mbox{.}(2011){Debuhr}, {Quataert}, \& {Ma}}]{DQM11}
{Debuhr} J., {Quataert} E., {Ma} C.-P., 2011, \mnras, 412, 1341, \ (DQM11)

\bibitem[{{Di Matteo} {et~al}\mbox{.}(2005){Di Matteo}, {Springel}, \&
  {Hernquist}}]{DSH05}
{Di Matteo} T., {Springel} V., {Hernquist} L., 2005, \nat, 433, 604

\bibitem[{{Dubois} {et~al}\mbox{.}(2012){Dubois}, {Devriendt}, {Slyz}, \&
  {Teyssier}}]{DDST12}
{Dubois} Y., {Devriendt} J., {Slyz} A., {Teyssier} R., 2012, \mnras, 420, 2662

\bibitem[{{Escala} {et~al}\mbox{.}(2004){Escala}, {Larson}, {Coppi}, \&
  {Mardones}}]{ELCM04}
{Escala} A., {Larson} R.~B., {Coppi} P.~S., {Mardones} D., 2004, \apj, 607, 765

\bibitem[{{Fabian}(1999{\natexlab{a}})}]{F99}
{Fabian} A.~C., 1999{\natexlab{a}}, Proceedings of the National Academy of
  Science, 96, 4749

\bibitem[{{Fabian}(1999{\natexlab{b}})}]{F99a}
{Fabian} A.~C., 1999{\natexlab{b}}, \mnras, 308, L39

\bibitem[{{Fabian}(2012)}]{F12}
{Fabian} A.~C., 2012, \araa, 50, 455

\bibitem[{{Ferrarese} \& {Merritt}(2000)}]{FM00}
{Ferrarese} L., {Merritt} D., 2000, \apjl, 539, L9

\bibitem[{{Fioc} \& {Rocca-Volmerange}(1997)}]{FR97}
{Fioc} M., {Rocca-Volmerange} B., 1997, \aap, 326, 950

\bibitem[{{Frank} {et~al}\mbox{.}(2002){Frank}, {King}, \& {Raine}}]{FKR02}
{Frank} J., {King} A., {Raine} D.~J., 2002, {Accretion Power in Astrophysics:
  Third Edition}

\bibitem[{{Gebhardt} {et~al}\mbox{.}(2000){Gebhardt}, {Bender}, {Bower},
  {Dressler}, {Faber}, {Filippenko}, {Green}, {Grillmair}, {Ho}, {Kormendy},
  {Lauer}, {Magorrian}, {Pinkney}, {Richstone}, \& {Tremaine}}]{Getal00}
{Gebhardt} K. {et~al.}, 2000, \apjl, 539, L13

\bibitem[{{Gerritsen} \& {Icke}(1997)}]{GI97}
{Gerritsen} J.~P.~E., {Icke} V., 1997, \aap, 325, 972

\bibitem[{{Gillessen} {et~al}\mbox{.}(2009){Gillessen}, {Eisenhauer}, {Trippe},
  {Alexander}, {Genzel}, {Martins}, \& {Ott}}]{GETetal09}
{Gillessen} S., {Eisenhauer} F., {Trippe} S., {Alexander} T., {Genzel} R.,
  {Martins} F., {Ott} T., 2009, \apj, 692, 1075

\bibitem[{{Gingold} \& {Monaghan}(1977)}]{GM77}
{Gingold} R.~A., {Monaghan} J.~J., 1977, \mnras, 181, 375

\bibitem[{{Governato} {et~al}\mbox{.}(2007){Governato}, {Willman}, {Mayer},
  {Brooks}, {Stinson}, {Valenzuela}, {Wadsley}, \& {Quinn}}]{Getal07}
{Governato} F., {Willman} B., {Mayer} L., {Brooks} A., {Stinson} G.,
  {Valenzuela} O., {Wadsley} J., {Quinn} T., 2007, \mnras, 374, 1479

\bibitem[{{G{\"u}ltekin} {et~al}\mbox{.}(2009){G{\"u}ltekin}, {Richstone},
  {Gebhardt}, {Lauer}, {Tremaine}, {Aller}, {Bender}, {Dressler}, {Faber},
  {Filippenko}, {Green}, {Ho}, {Kormendy}, {Magorrian}, {Pinkney}, \&
  {Siopis}}]{Getal09}
{G{\"u}ltekin} K. {et~al.}, 2009, \apj, 698, 198

\bibitem[{{Hartigan} {et~al}\mbox{.}(1995){Hartigan}, {Edwards}, \&
  {Ghandour}}]{HEG95}
{Hartigan} P., {Edwards} S., {Ghandour} L., 1995, \apj, 452, 736

\bibitem[{{Hobbs} {et~al}\mbox{.}(2012){Hobbs}, {Power}, {Nayakshin}, \&
  {King}}]{HPNK12}
{Hobbs} A., {Power} C., {Nayakshin} S., {King} A.~R., 2012, \mnras, 421, 3443

\bibitem[{{Hopkins} {et~al}\mbox{.}(2009){Hopkins}, {Cox}, {Younger}, \&
  {Hernquist}}]{HCYL09}
{Hopkins} P.~F., {Cox} T.~J., {Younger} J.~D., {Hernquist} L., 2009, \apj, 691,
  1168

\bibitem[{{Hopkins} \& {Quataert}(2010)}]{HQ10}
{Hopkins} P.~F., {Quataert} E., 2010, \mnras, 407, 1529

\bibitem[{{Johansson} {et~al}\mbox{.}(2009){Johansson}, {Naab}, \&
  {Burkert}}]{JNB09}
{Johansson} P.~H., {Naab} T., {Burkert} A., 2009, \apj, 690, 802

\bibitem[{{Katz}(1992)}]{k92}
{Katz} N., 1992, \apj, 391, 502

\bibitem[{{Kaufmann} {et~al}\mbox{.}(2007){Kaufmann}, {Mayer}, {Wadsley},
  {Stadel}, \& {Moore}}]{KMWSM07}
{Kaufmann} T., {Mayer} L., {Wadsley} J., {Stadel} J., {Moore} B., 2007, \mnras,
  375, 53

\bibitem[{{Kawakatu} \& {Umemura}(2002)}]{KU02}
{Kawakatu} N., {Umemura} M., 2002, \mnras, 329, 572

\bibitem[{{Kawata} \& {Gibson}(2005)}]{KG05}
{Kawata} D., {Gibson} B.~K., 2005, \mnras, 358, L16

\bibitem[{{Kennicutt}(1998)}]{K98}
{Kennicutt}, Jr. R.~C., 1998, \apj, 498, 541

\bibitem[{{Khan} {et~al}\mbox{.}(2011){Khan}, {Just}, \& {Merritt}}]{KJM11}
{Khan} F.~M., {Just} A., {Merritt} D., 2011, \apj, 732, 89

\bibitem[{{King}(2003)}]{K03}
{King} A., 2003, \apjl, 596, L27

\bibitem[{{Kormendy} \& {Richstone}(1995)}]{KR95}
{Kormendy} J., {Richstone} D., 1995, \araa, 33, 581

\bibitem[{{Kuijken} \& {Dubinski}(1995)}]{KD95}
{Kuijken} K., {Dubinski} J., 1995, \mnras, 277, 1341

\bibitem[{{Kurosawa} {et~al}\mbox{.}(2009){Kurosawa}, {Proga}, \&
  {Nagamine}}]{KPN09}
{Kurosawa} R., {Proga} D., {Nagamine} K., 2009, \apj, 707, 823

\bibitem[{{Lucy}(1977)}]{L77}
{Lucy} L.~B., 1977, \aj, 82, 1013

\bibitem[{{Madau} {et~al}\mbox{.}(1996){Madau}, {Ferguson}, {Dickinson},
  {Giavalisco}, {Steidel}, \& {Fruchter}}]{MFSGSF96}
{Madau} P., {Ferguson} H.~C., {Dickinson} M.~E., {Giavalisco} M., {Steidel}
  C.~C., {Fruchter} A., 1996, \mnras, 283, 1388

\bibitem[{{Magorrian} {et~al}\mbox{.}(1998){Magorrian}, {Tremaine},
  {Richstone}, {Bender}, {Bower}, {Dressler}, {Faber}, {Gebhardt}, {Green},
  {Grillmair}, {Kormendy}, \& {Lauer}}]{Metal98}
{Magorrian} J. {et~al.}, 1998, \aj, 115, 2285

\bibitem[{{Marconi} \& {Hunt}(2003)}]{MH03}
{Marconi} A., {Hunt} L.~K., 2003, \apjl, 589, L21

\bibitem[{{McLure} \& {Dunlop}(2002)}]{MD02}
{McLure} R.~J., {Dunlop} J.~S., 2002, \mnras, 331, 795

\bibitem[{{McNamara} \& {Nulsen}(2007)}]{MN07}
{McNamara} B.~R., {Nulsen} P.~E.~J., 2007, \araa, 45, 117

\bibitem[{{McNamara} {et~al}\mbox{.}(2005){McNamara}, {Nulsen}, {Wise},
  {Rafferty}, {Carilli}, {Sarazin}, \& {Blanton}}]{MNetal05}
{McNamara} B.~R., {Nulsen} P.~E.~J., {Wise} M.~W., {Rafferty} D.~A., {Carilli}
  C., {Sarazin} C.~L., {Blanton} E.~L., 2005, \nat, 433, 45

\bibitem[{{McNamara} {et~al}\mbox{.}(2000){McNamara}, {Wise}, {Nulsen},
  {David}, {Sarazin}, {Bautz}, {Markevitch}, {Vikhlinin}, {Forman}, {Jones}, \&
  {Harris}}]{Metal00}
{McNamara} B.~R. {et~al.}, 2000, \apjl, 534, L135

\bibitem[{{Meier}(2001)}]{M01}
{Meier} D.~L., 2001, \apjl, 548, L9

\bibitem[{{Mihos} \& {Hernquist}(1996)}]{MH96}
{Mihos} J.~C., {Hernquist} L., 1996, \apj, 464, 641

\bibitem[{{Mori} {et~al}\mbox{.}(1997){Mori}, {Yoshii}, {Tsujimoto}, \&
  {Nomoto}}]{MYTN97}
{Mori} M., {Yoshii} Y., {Tsujimoto} T., {Nomoto} K., 1997, \apjl, 478, L21

\bibitem[{{Moster} {et~al}\mbox{.}(2011){Moster}, {Macci{\`o}}, {Somerville},
  {Naab}, \& {Cox}}]{MMSNC11}
{Moster} B.~P., {Macci{\`o}} A.~V., {Somerville} R.~S., {Naab} T., {Cox} T.~J.,
  2011, \mnras, 415, 3750

\bibitem[{{Murray} {et~al}\mbox{.}(2005){Murray}, {Quataert}, \&
  {Thompson}}]{MQT05}
{Murray} N., {Quataert} E., {Thompson} T.~A., 2005, \apj, 618, 569

\bibitem[{{Narayan} {et~al}\mbox{.}(1998){Narayan}, {Mahadevan}, \&
  {Quataert}}]{NMQ98}
{Narayan} R., {Mahadevan} R., {Quataert} E., 1998, in {Abramowicz} M.~A.,
  {Bjornsson} G., {Pringle} J.~E., eds, Theory of Black Hole Accretion Disks.
  Cambridge Univ. Press, Cambridge, p. 148

\bibitem[{{Navarro} \& {White}(1993)}]{NW93}
{Navarro} J.~F., {White} S.~D.~M., 1993, \mnras, 265, 271

\bibitem[{{Nobukawa} {et~al}\mbox{.}(2011){Nobukawa}, {Ryu}, {Tsuru}, \&
  {Koyama}}]{NRTK11}
{Nobukawa} M., {Ryu} S.~G., {Tsuru} T.~G., {Koyama} K., 2011, \apjl, 739, L52

\bibitem[{{Okamoto} {et~al}\mbox{.}(2008){Okamoto}, {Nemmen}, \&
  {Bower}}]{ONB08}
{Okamoto} T., {Nemmen} R.~S., {Bower} R.~G., 2008, \mnras, 385, 161, \ (ONB08)

\bibitem[{{Power} {et~al}\mbox{.}(2011){Power}, {Nayakshin}, \& {King}}]{PNK11}
{Power} C., {Nayakshin} S., {King} A., 2011, \mnras, 412, 269

\bibitem[{{Proga} {et~al}\mbox{.}(2000){Proga}, {Stone}, \& {Kallman}}]{PSK00}
{Proga} D., {Stone} J.~M., {Kallman} T.~R., 2000, \apj, 543, 686

\bibitem[{{Rasmussen} {et~al}\mbox{.}(2009){Rasmussen}, {Sommer-Larsen},
  {Pedersen}, {Toft}, {Benson}, {Bower}, \& {Grove}}]{Retal09}
{Rasmussen} J., {Sommer-Larsen} J., {Pedersen} K., {Toft} S., {Benson} A.,
  {Bower} R.~G., {Grove} L.~F., 2009, \apj, 697, 79

\bibitem[{{Reynolds}(2008)}]{R08}
{Reynolds} S.~P., 2008, \araa, 46, 89

\bibitem[{{Robertson} {et~al}\mbox{.}(2006){Robertson}, {Hernquist}, {Cox}, {Di
  Matteo}, {Hopkins}, {Martini}, \& {Springel}}]{RHCDHMS06}
{Robertson} B., {Hernquist} L., {Cox} T.~J., {Di Matteo} T., {Hopkins} P.~F.,
  {Martini} P., {Springel} V., 2006, \apj, 641, 90

\bibitem[{{Sanders} {et~al}\mbox{.}(1988){Sanders}, {Soifer}, {Elias},
  {Madore}, {Matthews}, {Neugebauer}, \& {Scoville}}]{Setal88}
{Sanders} D.~B., {Soifer} B.~T., {Elias} J.~H., {Madore} B.~F., {Matthews} K.,
  {Neugebauer} G., {Scoville} N.~Z., 1988, \apj, 325, 74

\bibitem[{{Sarajedini} {et~al}\mbox{.}(2011){Sarajedini}, {Koo}, {Klesman},
  {Laird}, {Perez Gonzalez}, \& {Mozena}}]{SKKLPM11}
{Sarajedini} V.~L., {Koo} D.~C., {Klesman} A.~J., {Laird} E.~S., {Perez
  Gonzalez} P.~G., {Mozena} M., 2011, \apj, 731, 97

\bibitem[{{Scannapieco} {et~al}\mbox{.}(2012){Scannapieco}, {Wadepuhl},
  {Parry}, {Navarro}, {Jenkins}, {Springel}, {Teyssier}, {Carlson}, {Couchman},
  {Crain}, {Vecchia}, {Frenk}, {Kobayashi}, {Monaco}, {Murante}, {Okamoto},
  {Quinn}, {Schaye}, {Stinson}, {Theuns}, {Wadsley}, {White}, \&
  {Woods}}]{Setal12}
{Scannapieco} C. {et~al.}, 2012, \mnras, 423, 1726

\bibitem[{{Scannapieco} \& {Oh}(2004)}]{SO04}
{Scannapieco} E., {Oh} S.~P., 2004, \apj, 608, 62

\bibitem[{{Scannapieco} {et~al}\mbox{.}(2005){Scannapieco}, {Silk}, \&
  {Bouwens}}]{SSB05}
{Scannapieco} E., {Silk} J., {Bouwens} R., 2005, \apjl, 635, L13

\bibitem[{{Shakura} \& {Sunyaev}(1973)}]{SS73}
{Shakura} N.~I., {Sunyaev} R.~A., 1973, \aap, 24, 337

\bibitem[{{Shaver} {et~al}\mbox{.}(1996){Shaver}, {Wall}, {Kellermann},
  {Jackson}, \& {Hawkins}}]{SWKJH96}
{Shaver} P.~A., {Wall} J.~V., {Kellermann} K.~I., {Jackson} C.~A., {Hawkins}
  M.~R.~S., 1996, \nat, 384, 439

\bibitem[{{Sijacki} {et~al}\mbox{.}(2007){Sijacki}, {Springel}, {Di Matteo}, \&
  {Hernquist}}]{SSDH07}
{Sijacki} D., {Springel} V., {Di Matteo} T., {Hernquist} L., 2007, \mnras, 380,
  877

\bibitem[{{Silk} \& {Rees}(1998)}]{SR98}
{Silk} J., {Rees} M.~J., 1998, \aap, 331, L1

\bibitem[{{Sommer-Larsen} {et~al}\mbox{.}(1999){Sommer-Larsen}, {Gelato}, \&
  {Vedel}}]{SGV99}
{Sommer-Larsen} J., {Gelato} S., {Vedel} H., 1999, \apj, 519, 501

\bibitem[{{Springel} {et~al}\mbox{.}(2005){Springel}, {Di Matteo}, \&
  {Hernquist}}]{SDH05}
{Springel} V., {Di Matteo} T., {Hernquist} L., 2005, \mnras, 361, 776, \
  (SDH05)

\bibitem[{{Springel} \& {Hernquist}(2002)}]{SH02}
{Springel} V., {Hernquist} L., 2002, \mnras, 333, 649

\bibitem[{{Springel} \& {Hernquist}(2003)}]{SH03}
{Springel} V., {Hernquist} L., 2003, \mnras, 339, 289

\bibitem[{{Springel} {et~al}\mbox{.}(2001){Springel}, {Yoshida}, \&
  {White}}]{SYW01}
{Springel} V., {Yoshida} N., {White} S.~D.~M., 2001, \na, 6, 79

\bibitem[{{Stinson} {et~al}\mbox{.}(2006){Stinson}, {Seth}, {Katz}, {Wadsley},
  {Governato}, \& {Quinn}}]{SSKWGQ06}
{Stinson} G., {Seth} A., {Katz} N., {Wadsley} J., {Governato} F., {Quinn} T.,
  2006, \mnras, 373, 1074

\bibitem[{{Thacker} \& {Couchman}(2000)}]{TC00}
{Thacker} R.~J., {Couchman} H.~M.~P., 2000, \apj, 545, 728

\bibitem[{{Thacker} \& {Couchman}(2006)}]{TC06}
{Thacker} R.~J., {Couchman} H.~M.~P., 2006, Computer Physics Communications,
  174, 540

\bibitem[{{Thacker} {et~al}\mbox{.}(2006){Thacker}, {Scannapieco}, \&
  {Couchman}}]{TSC06}
{Thacker} R.~J., {Scannapieco} E., {Couchman} H.~M.~P., 2006, \apj, 653, 86

\bibitem[{{Tremaine} {et~al}\mbox{.}(2002){Tremaine}, {Gebhardt}, {Bender},
  {Bower}, {Dressler}, {Faber}, {Filippenko}, {Green}, {Grillmair}, {Ho},
  {Kormendy}, {Lauer}, {Magorrian}, {Pinkney}, \& {Richstone}}]{Tetal02}
{Tremaine} S. {et~al.}, 2002, \apj, 574, 740

\bibitem[{{Umemura}(2001)}]{U01}
{Umemura} M., 2001, \apjl, 560, L29

\bibitem[{{Umemura} {et~al}\mbox{.}(1997){Umemura}, {Fukue}, \&
  {Mineshige}}]{UFM97}
{Umemura} M., {Fukue} J., {Mineshige} S., 1997, \apjl, 479, L97

\bibitem[{{Webb} \& {Malkan}(2000)}]{WM00}
{Webb} W., {Malkan} M., 2000, \apj, 540, 652

\bibitem[{{Widrow} \& {Dubinski}(2005)}]{WD05}
{Widrow} L.~M., {Dubinski} J., 2005, \apj, 631, 838

\bibitem[{{Widrow} {et~al}\mbox{.}(2008){Widrow}, {Pym}, \& {Dubinski}}]{WPD08}
{Widrow} L.~M., {Pym} B., {Dubinski} J., 2008, \apj, 679, 1239

\bibitem[{{Williamson} \& {Thacker}(2012)}]{WT12}
{Williamson} D.~J., {Thacker} R.~J., 2012, \mnras, 421, 2170

\bibitem[{{Willson}(2000)}]{W00}
{Willson} L.~A., 2000, \araa, 38, 573

\bibitem[{{Wurster} \& {Thacker}(2013)}]{WT13power}
{Wurster} J., {Thacker} R.~J., 2013, ArXiv e-prints

\end{thebibliography}

\label{lastpage}
\end{document}